\documentclass[usenatbib]{mnras}

\usepackage{newtxtext,newtxmath}
\usepackage[T1]{fontenc}
\usepackage{ae,aecompl}

\usepackage{graphicx}	% Including figure files
\usepackage{amsmath}	% Advanced maths commands
\usepackage[usenames, dvipsnames]{color}
\usepackage{url}

\title[Massive Galaxy Main Sequence]{The Shape and Scatter of The Galaxy Main Sequence for Massive Galaxies at Cosmic Noon}

\author[Sherman et al.]{Sydney Sherman$^{1}$\thanks{E-mail: \texttt{sydney.sherman@utexas.edu}},
Shardha Jogee$^{1}$,
Jonathan Florez$^{1}$,
Steven L. Finkelstein$^{1}$, \newauthor
Robin Ciardullo$^{2,3}$, 
Isak Wold$^{4}$,
Matthew L. Stevans$^{1}$, \newauthor
Lalitwadee Kawinwanichakij$^{5,6,7}$, 
Casey Papovich$^{6,7}$, 
Caryl Gronwall$^{2,3}$
\\
$^{1}$Department of Astronomy, The University of Texas at Austin, Austin, TX 78712 \\
$^{2}$Department of Astronomy and Astrophysics, The Pennsylvania State University, University Park, PA 16802\\
$^{3}$Institute for Gravitation and the Cosmos, The Pennsylvania State University, University Park, PA 16802\\
$^{4}$NASA Goddard Space Flight Center, Greenbelt, MD 20771\\
$^{5}$Kavli Institute for the Physics and Mathematics of the Universe, The University of Tokyo, Kashiwa, Japan 277-8583 (Kavli IPMU, WPI)\\
$^{6}$Department of Physics and Astronomy, Texas A\&M University, College Station, TX, 77843-4242\\
$^{7}$George P. and Cynthia Woods Mitchell Institute for Fundamental Physics and Astronomy, Texas A\&M University, College Station, TX 77843\\
}

\date{Accepted 2021 May 5. Received 2021 April 10; in original form 2021 February 12}

\pubyear{2020}

\begin{document}
\label{firstpage}
\pagerange{\pageref{firstpage}--\pageref{lastpage}}
\maketitle

%%%%%%%%%%%%%%%%%%%%%%%%%%%%%%%%%%%%%%%%%%%%%%%%%%
%%%%%%%%%%%%%%%%% ABSTRACT %%%%%%%%%%%%%%%%%%
%%%%%%%%%%%%%%%%%%%%%%%%%%%%%%%%%%%%%%%%%%%%%%%%%%
\begin{abstract}
We present the main sequence for all galaxies and star-forming galaxies for a sample of 28,469 massive ($M_\star \ge 10^{11}$M$_\odot$) galaxies at cosmic noon ($1.5 < z < 3.0$), uniformly selected from a 17.5 deg$^2$ area (0.33 Gpc$^3$ comoving volume at these redshifts). Our large sample allows for a novel approach to investigating the galaxy main sequence that has not been accessible to previous studies. We measure the main sequence in small mass bins in the SFR-M$_{\star}$ plane without assuming a functional form for the main sequence. With a large sample of galaxies in each mass bin, we isolate star-forming galaxies by locating the transition between the star-forming and green valley populations in the SFR-M$_{\star}$ plane. This approach eliminates the need for arbitrarily defined fixed cutoffs when isolating the star-forming galaxy population, which often biases measurements of the scatter around the star-forming galaxy main sequence. We find that the main sequence for all galaxies becomes increasingly flat towards present day at the high-mass end, while the star-forming galaxy main sequence does not. We attribute this difference to the increasing fraction of the collective green valley and quiescent galaxy population from $z=3.0$ to $z=1.5$. Additionally, we measure the total scatter around the star-forming galaxy main sequence and find that it is $\sim0.5-1.0$ dex with little evolution as a function of mass or redshift. We discuss the implications that these results have for pinpointing the physical processes driving massive galaxy evolution.
\end{abstract}


%%%%%%%%%%%%%%%%%%%%%%%%%%%%%%%%%%%%%%%%%%%%%%%%%%
%%%%%%%%%%%%%%%%% BODY OF PAPER %%%%%%%%%%%%%%%%%%
%%%%%%%%%%%%%%%%%%%%%%%%%%%%%%%%%%%%%%%%%%%%%%%%%%

%%%%%%%%%%%%%%%%%%%%%%%%%%%%%%%%%%%%%%%%%%%%%%%%%%
%%%%%%%%%%%%%%%%% INTRODUCTION %%%%%%%%%%%%%%%%%%
%%%%%%%%%%%%%%%%%%%%%%%%%%%%%%%%%%%%%%%%%%%%%%%%%%
\section{Introduction}
%\paragraph*{Big Question on galaxy evolution}
The way in which massive galaxies build their stellar populations, and achieve this earlier than lower mass populations, remains an important question in the study of galaxy evolution. Theoretical models (e.g., \citealt{Somerville1999}, \citealt{Cole2000}, \citealt{Bower2006}, \citealt{Croton2006}, \citealt{Somerville2008}, \citealt{Benson2012}, \citealt{Somerville2015} and references therein, \citealt{Croton2016}, \citealt{Naab2017} and references therein, \citealt{Weinberger2017}, \citealt{Cora2018}, \citealt{Knebe2018}, \citealt{Behroozi2019}, \citealt{Cora2019}, \citealt{Dave2019}) struggle to implement physical processes that can simultaneously reproduce the observed properties of the massive and low mass galaxy populations at both high and low redshifts (e.g., \citealt{Conselice2007}, \citealt{Asquith2018}, \citealt{Sherman2020}, \citealt{Sherman2020b}). Observations of large samples of massive galaxies at cosmic noon ($1.5 < z < 3.0$), a time when the massive galaxy population transitions from star-forming to quiescent (e.g., \citealt{Conselice2011}, \citealt{vanderWel2011}. \citealt{Weinzirl2011},  \citealt{Muzzin2013}, \citealt{vanDokkum2015}, \citealt{Martis2016}, \citealt{Tomczak2016}, \citealt{Sherman2020b}), can provide important constraints on the physical processes driving the early assembly of massive galaxies.

%\paragraph*{Observables (e.g. SMF or SFR-M*) to help answer big questions}
The stellar masses and star-formation rates of galaxies at cosmic noon ($1.5 < z < 3.0$) are fundamental quantities that provide insights into this dynamic period in the history of the universe. At this epoch, proto-clusters began to collapse into the rich clusters seen at present day (e.g., \citealt{Gobat2011}, 
\citealt{Lotz2013}, \citealt{Overzier2016}, \citealt{Wang2016}, \citealt{Chiang2017}), star-formation and black hole accretion peaked (e.g., \citealt{MadauDickinson2014}), and the massive ($M_\star \ge 10^{11}$M$_\odot$) galaxy population transitioned from being predominantly star-forming to predominantly quiescent (e.g., \citealt{Conselice2011}, \citealt{vanderWel2011}. \citealt{Weinzirl2011},  \citealt{Muzzin2013}, \citealt{vanDokkum2015}, \citealt{Martis2016}, \citealt{Tomczak2016}, \citealt{Sherman2020b}). The relationship between star-formation rate and stellar mass, coined the ``main sequence" by \cite{Noeske2007}, provides key insights into the formation history of the massive galaxy population. 

Although a significant number of studies (e.g., \citealt{Daddi2007}, \citealt{Elbaz2007}, \citealt{Noeske2007}, \citealt{Karim2011}, \citealt{Rodighiero2011}, \citealt{KGuo2013}, \citealt{Speagle2014}, \citealt{Whitaker2014}, \citealt{Lee2015}, \citealt{Renzini2015}, \citealt{Salmon2015}, \citealt{Schreiber2015}, \citealt{Tasca2015}, \citealt{Tomczak2016}, \citealt{Santini2017}, \citealt{Popesso2019}, among others) have investigated the nature of galaxies in the SFR-$M_\star$ plane, a consensus has not yet been reached for a single definition of the ``main sequence", specifically as it pertains to the star-forming galaxy main sequence. Some studies choose to pre-select for star-forming galaxies (e.g., \citealt{Noeske2007}, \citealt{Daddi2007}, \citealt{Whitaker2014}, \citealt{Tomczak2016}), typically via emission at 24$\mu$m. Others select a sample of star-forming galaxies from a sample containing all galaxies (e.g., \citealt{Whitaker2014} and \citealt{Tomczak2016} at intermediate redshifts), with techniques such as color-color selection. In this work, we investigate both the main sequence for all galaxies and for star-forming galaxies using a sample of massive ($M_\star \ge 10^{11}$M$_\odot$) galaxies that spans a wide range of specific star-formation rates in the star-forming, green valley (e.g., \citealt{Martin2007}, \citealt{Salim2007}, \citealt{Wyder2007}), and quiescent populations. 

%\paragraph*{Results by others on big questions and observables}
Previous studies have focused on three key aspects of the main sequence: the slope, normalization, and scatter around the main sequence. The slope provides information about when galaxies of different masses begin to quench (the so-called ``downsizing" scenario; \citealt{Cowie1996}). Out to $z\sim6$ the power-law slope is measured to be between $\sim0-1$ (e.g., \citealt{Daddi2007}, \citealt{Elbaz2007}, \citealt{Noeske2007}, \citealt{Rodighiero2011}, \citealt{KGuo2013}, \citealt{Speagle2014}, \citealt{Whitaker2014}, \citealt{Renzini2015}, \citealt{Tomczak2016}, \citealt{Santini2017}, among others), with evidence that higher mass star-forming galaxies exhibit a shallower slope than lower mass star-forming galaxies (e.g., \citealt{Whitaker2014}, \citealt{Lee2015}, \citealt{Tasca2015}). The normalization of the main sequence has been shown to increase with increasing redshift, indicating that the specific star-formation rates of extreme galaxies found at late times were more typical specific star-formation rates at earlier times (e.g., \citealt{Karim2011}, \citealt{Speagle2014}, \citealt{Whitaker2014}, \citealt{Tomczak2016}, \citealt{Santini2017}). Finally, the star-forming galaxy main sequence relation has been found to be quite tight with a rather constant scatter (typical scatter is measured to be $0.2 - 0.4$ dex; e.g., \citealt{Rodighiero2011}, \citealt{Speagle2014}, \citealt{Schreiber2015}, \citealt{Popesso2019}) with the level of scatter often attributed to the level of stochasticity in the star-formation history of the population (e.g., \citealt{Caplar2019},  \citealt{Matthee2019}). 

%\paragraph*{Techniques/challenges in getting these observables + limitations of  earlier work}
Typically, previous studies have focused on achieving deep observations taken over small areas, often pushing constraints of the main sequence to fairly low masses ($\sim10^8 - 10^9$M$_\odot$; e.g., \citealt{Whitaker2014}, \citealt{Tomczak2016}). Different methods taken by previous studies for measuring the main sequence (e.g., extrapolation from low to high masses, fitting single and double power laws, stacking analyses, etc.), as well as inconsistent (and often biased) methods of separating star-forming galaxies from the total population (e.g., color-color indicators, specific star-formation rate thresholds, distance below the main sequence, detection in particular filters, etc.) have led to measures of the main sequence that are not unbiased \citep{Renzini2015}. Furthermore, biased selection of star-forming galaxies has often forced previous works to make assumptions about the distribution of star-forming galaxies in the SFR-M$_{\star}$ plane, which makes robust measures of the scatter around the star-forming galaxy main sequence, a strong tracer of the stochasticity of star-formation histories, quite difficult. 

%\paragraph*{What your work adds to a) and d)}
In this work, we present the massive end of the galaxy main sequence for all galaxies and star-forming galaxies at cosmic noon ($1.5 < z < 3.0$) using a sample of 28,469 massive ($M_\star \ge 10^{11}$M$_\odot$) galaxies. Notably, we do not make any assumptions about the functional form of the galaxy main sequence nor do we make assumptions about the distribution of massive galaxies in the SFR-M$_{\star}$ plane. This novel, unbiased approach is made possible by our large sample which is uniformly selected from a 17.5 deg$^2$ area ($\sim0.33$ Gpc$^3$ comoving volume over $1.5 < z < 3.0$), significantly reducing Poisson errors and rendering the effects of cosmic variance negligible. With this large sample, we are uniquely suited to separate star-forming galaxies from the collective green valley and quiescent galaxy populations by locating the transition between the star-forming and green valley populations in the SFR-M$_{\star}$ plane in small mass bins, rather than using fixed cutoffs to define these populations. Finally, due to our meaningful separation of star-forming galaxies from the total population, we are able to perform an unbiased study of the scatter around the star-forming galaxy main sequence as a function of stellar mass. 

We also compare our empirical results with those from hydrodynamical models SIMBA \citep{Dave2019} and IllustrisTNG (\citealt{Pillepich2018b}, \citealt{Springel2018}, \citealt{Nelson2018}, \citealt{Naiman2018}, \citealt{Marinacci2018}), as well as the semi-analytic model SAG \citep{Cora2018}. \cite{Sherman2020b} showed that these models face significant challenges in reproducing the observed quiescent fraction of massive ($M_\star \ge 10^{11}$M$_\odot$) galaxies at $1.5 < z < 3.0$, indicating that the implementation of the physical processes underlying massive galaxy evolution at these epochs may need to be revised. 

%\paragraph*{Organization of paper}
This paper is organized as follows. In Section \ref{sec:data_and_analysis} we detail the data used in this work, the SED fitting procedure, and sample selection. Section \ref{sec:MainSequence} presents our measurement of the main sequence for all galaxies, Section \ref{sec:SF_MainSequence} presents the main sequence for star-forming galaxies, and in Section \ref{sec:MSAll_vs_MSSF} we compare the resulting main sequences for all and star-forming galaxies. In Section \ref{sec:MS_scatter} we measure the scatter around the star-forming galaxy main sequence. In Section \ref{sec:CompWObs} we present comparisons with previous observational studies, and in Section \ref{sec:CompWTheory} we compare our empirical result with those from theoretical models. Finally, we discuss the implications of our results in Section \ref{sec:Discussion} and summarize our results in Section \ref{sec:Summary}. Throughout this work we adopt a flat $\Lambda$CDM cosmology with $h = 0.7$, $\Omega_m = 0.3$, and $\Omega_{\Lambda} = 0.7$. 

%%%%%%%%%%%%%%%%%%%%%%%%%%%%%%%%%%%%%%%%%%%%%%%%%%
%%%%%%%%%%%%%%%%% DATA %%%%%%%%%%%%%%%%%%
%%%%%%%%%%%%%%%%%%%%%%%%%%%%%%%%%%%%%%%%%%%%%%%%%%
\section{Data and Analysis}
\label{sec:data_and_analysis}
The data, SED fitting, sample selection, and stellar mass completeness estimates used in this work are the same as those used in \cite{Sherman2020b} and will briefly be described here. Our catalog is NEWFIRM $K_s$-selected (depth 22.4 AB mag at 5$\sigma$; PI Finkelstein, \citealt{Stevans2021}) and covers 17.5 deg$^2$ in the SDSS Stripe 82 equatorial field. In addition to the NEWFIRM $K_s$ data, we also utilize \textit{u, g, r, i, z}  photometry from the Dark Energy Camera (DECam) (\citealt{Wold2019}, \citealt{Kawinwanichakij2020}, \citealt{Stevans2021}; r-band 5$\sigma$ depth is r = 24.5 AB mag), VICS82 $J$ and $K_s$ data (\citealt{Geach2017}; 5$\sigma$ depth for J-band is 21.5 AB mag and for K-band is 20.9 AB mag), and \textit{Spitzer}-IRAC 3.6 and 4.5$\mu$m photometry (PI Papovich; \citealt{Papovich2016}, \cite{Kawinwanichakij2020}); 5$\sigma$ depth is 22 AB mag in both filters). Combined, these data provide up to 10 photometric data points with which we can use SED fitting techniques to estimate redshift, stellar mass, SFR, and other galaxy properties. Additional photometric data in this footprint (which are not used in SED fitting) come from \textit{Herschel}-SPIRE (HerS, \citealt{Viero2014}) far-IR/submillimeter, and XMM-Newton and Chandra X-ray Observatory X-ray data from the Stripe 82X survey (\citealt{LaMassa2013a}, \citealt{LaMassa2013b}, \citealt{Ananna2017}, the X-ray data cover $\sim$11.2 deg$^2$). In this region, optical ($3500-5500$\AA) spectroscopy is being obtained by the Hobby Eberly Telescope Dark Energy Experiment (HETDEX, \citealt{Hill2008}), and these data are only used to estimate the accuracy of photometric redshifts, when available (see below). 

SED fitting is performed using EAZY-py\footnote{The version of EAZY-py used in this work was downloaded in May 2018 from \url{https://github.com/gbrammer/eazy-py} and was later modified by \cite{Sherman2020} (also described in \citealt{Sherman2020b}) to add functions that provide uncertainties on measured galaxy parameters.}, a python-based version of EAZY \citep{Brammer2008}, which simultaneously fits for photometric redshift, stellar mass, SFR, and other galaxy properties, with an implementation from \cite{Sherman2020} and \cite{Sherman2020b} that also gives error estimates for these parameters (finding typical stellar mass and SFR errors of $\pm0.08$ dex and $\pm0.18$ dex, respectively for $1.5 < z < 3.0$ galaxies above our estimated mass completeness limits, detailed below). EAZY-py performs SED fitting using twelve Flexible Stellar Population Synthesis (FSPS; \citealt{Conroy2009}, \citealt{Conroy2010}) templates in non-negative linear combination. Our SED fitting is performed using the default EAZY-py FSPS templates which are built with a \cite{Chabrier2003} initial mass function (IMF), the \cite{KriekConroy2013} dust law, solar metallicity, and star-formation histories including bursty and slowly rising models. 

We note that recent studies (e.g., \citealt{Carnall2019}, \citealt{Leja2019}) have showed the strong influence that the chosen star-formation history has on the resultant SFR given by SED fitting. In our study, the EAZY-py fitting method constructs a best-fit SED from the non-negative linear combination of twelve templates, each with different star-formation histories. Because of this, the resultant best-fit SED is not restricted to a single underlying star-formation history. Additionally, \cite{Sherman2020} used a diverse set of mock galaxies (V. Acquaviva, private communication) constructed with \cite{BruzualCharlot2003} templates and spanning stellar masses up to $M_{\star}=10^{12}$M$_{\odot}$ to validate the SED fitting procedure described above. These models were constructed from various underlying dust laws, IMFs, and star-formation histories (including exponentially declining, delayed exponential, constant, and linearly increasing). \cite{Sherman2020} found that for galaxies at $1.5 < z < 3.0$, EAZY-py is able to adequately recover the redshift, stellar mass, and SFR for the mock galaxies. 

Photometric redshift accuracy is estimated using spectroscopic redshifts from SDSS \citep{Eisenstein2011} at $z < 1$ and the second internal data release of the HETDEX survey \citep{Hill2008} at $1.9 < z < 3.5$. For both samples, \cite{Sherman2020b} quantified the photometric redshift recovery using the normalized median absolute deviation ($\sigma_{\rm NMAD}$; \citealt{Brammer2008}). Using the low-redshift sample from SDSS $\sigma_{\rm NMAD}$ = 0.053, and for the intermediate redshift galaxies from HETDEX $\sigma_{\rm NMAD}$ = 0.102. This intermediate redshift sample has only 56 galaxies, which are all visually inspected to confirm the spectroscopic redshift from the HETDEX pipeline, and this sample is expected to grow with future data releases. Three of the 56 intermediate redshift galaxies are catastrophic outliers (5.3\%) where the HETDEX spectrum places them at ($z < 0.5$) but the best-fit photometric redshift is $z > 2$. We note that catastrophic outliers are not removed from the low or high redshift samples before computing $\sigma_{\rm NMAD}$. 

Our science sample is the same as that from \cite{Sherman2020b}, comprised of 54,001 galaxies at $1.5 < z < 3.0$, of which, 28,469 are fit to have $M_\star \ge 10^{11}$M$_\odot$. The 95\% stellar mass completeness limits for this sample are log($M_{\star}$/$\rm M_{\odot}$) = 10.69, 10.86, and 11.13 in our $1.5 < z < 2.0$, $2.0 < z < 2.5$, and $2.5 < z < 3.0$ bins, respectively. We refer the reader to \cite{Sherman2020b} for details regarding sample selection and mass completeness estimates. 

For every galaxy in our $K_s$-selected sample, we obtain a measure of dust-corrected SFR, with an associated uncertainty, from our SED fitting procedure. The SED fitting procedure uses all available filters to find the best-fitting SED. Unlike the rather straightforward connection between a galaxy's $K_s$-band magnitude and that galaxy's stellar mass, there is not a straightforward connection between the measured dust-corrected SFR and a particular band. To estimate an SFR completeness, however, we can use the g-band as a proxy for FUV flux (see \citealt{Sherman2020b} and \citealt{Florez2020}) and obtain a g-band SFR completeness estimate. To achieve this, we take the $5\sigma$ g-band limiting magnitude for our survey ($\rm m_{\rm g,lim} = 24.8$ mag AB; computed by \citealt{Wold2019}) and, following \cite{Sherman2020b} and \cite{Florez2020}, we apply the conversion factor from \cite{Hao2011} to convert the  $5\sigma$ g-band limiting magnitude into an estimate of SFR$_{\rm FUV}$. The \cite{Hao2011} conversion assumes a \cite{Kroupa2001} IMF, and we reduce the estimated SFR$_{\rm FUV}$ by 0.046 dex to align the results with the \cite{Chabrier2003} IMF used throughout this work. We estimate the g-band based SFR completeness limits to be SFR = 2.36, 4.39, and 7.16 M$_{\odot}$ yr$^{-1}$ in our $1.5 < z < 2.0$, $2.0 < z < 2.5$, and $2.5 < z < 3.0$ bins, respectively. If we further apply a dust correction based on the median extinction measured by our SED fitting procedure for galaxies within $\pm0.1$ mag of the $5\sigma$ g-band completeness limit (this value is $\sim0.8-1.0$A$_{\rm v}$ across our three redshift bins), we find that the dust-corrected g-band SFR completeness limits are SFR = 4.80, 8.89, and 17.56 M$_{\odot}$ yr$^{-1}$ in our $1.5 < z < 2.0$, $2.0 < z < 2.5$, and $2.5 < z < 3.0$ bins, respectively.  Again, we emphasize that because our sample is $K_s$ selected, not g-band selected, for every object in our science sample we have a measurement, from SED fitting, of the dust-corrected SFR with an associated uncertainty. This holds true even for those with SFR measured by our SED fitting procedure to be below the $5\sigma$ g-band based SFR completeness limit estimates. 

%%%%%%%%%%%%%%%%%%%%%%%%%%%%%%%%%%%%%%%%%%%%%%%%%%
%%%%%%%%%%%%%%%%% RESULTS: SHELA ONLY %%%%%%%%%%%%%%%%%%
%%%%%%%%%%%%%%%%%%%%%%%%%%%%%%%%%%%%%%%%%%%%%%%%%%
\section{Galaxy Main Sequence} 
\label{sec:MainSequence_Opening}
In this Section we present the main sequence for all galaxies, which is computed in individual small mass bins at the high mass end, thereby eliminating the need for extrapolation or assumed functional forms. We also detail a novel method for isolating the star-forming galaxy population in an unbiased way, and we use this sample to explore the main sequence for star-forming galaxies. Finally we compare the main sequence for all galaxies with the main sequence for star-forming galaxies, and detail how the buildup of the collective green valley and quiescent galaxy populations influences the time evolution of the slope of the main sequences for all galaxies and star-forming galaxies. 

\subsection{Measuring the Main Sequence for All Galaxies} 
\label{sec:MainSequence}
The main sequence in each of our three redshift bins spanning $1.5 < z < 3.0$ is defined to be the average SFR in small mass bins in the SFR-M$_{\star}$ plane. To compute the error on the main sequence, we employ a bootstrap resampling procedure (\citealt{Sherman2020b}, \citealt{Florez2020}) that is repeated 1000 times. During each bootstrap draw we select a random sample of galaxies from each mass bin, with replacement, where the sample size is equal to the number of galaxies in the bin. By taking the average SFR in each of the 1000 draws, we generate a distribution of average SFR (main sequence) values. The lower and upper error bars on the main sequence are the 16th and 84th percentiles of this distribution, respectively. 

\begin{figure*}
\begin{center}
\includegraphics[width=\textwidth]{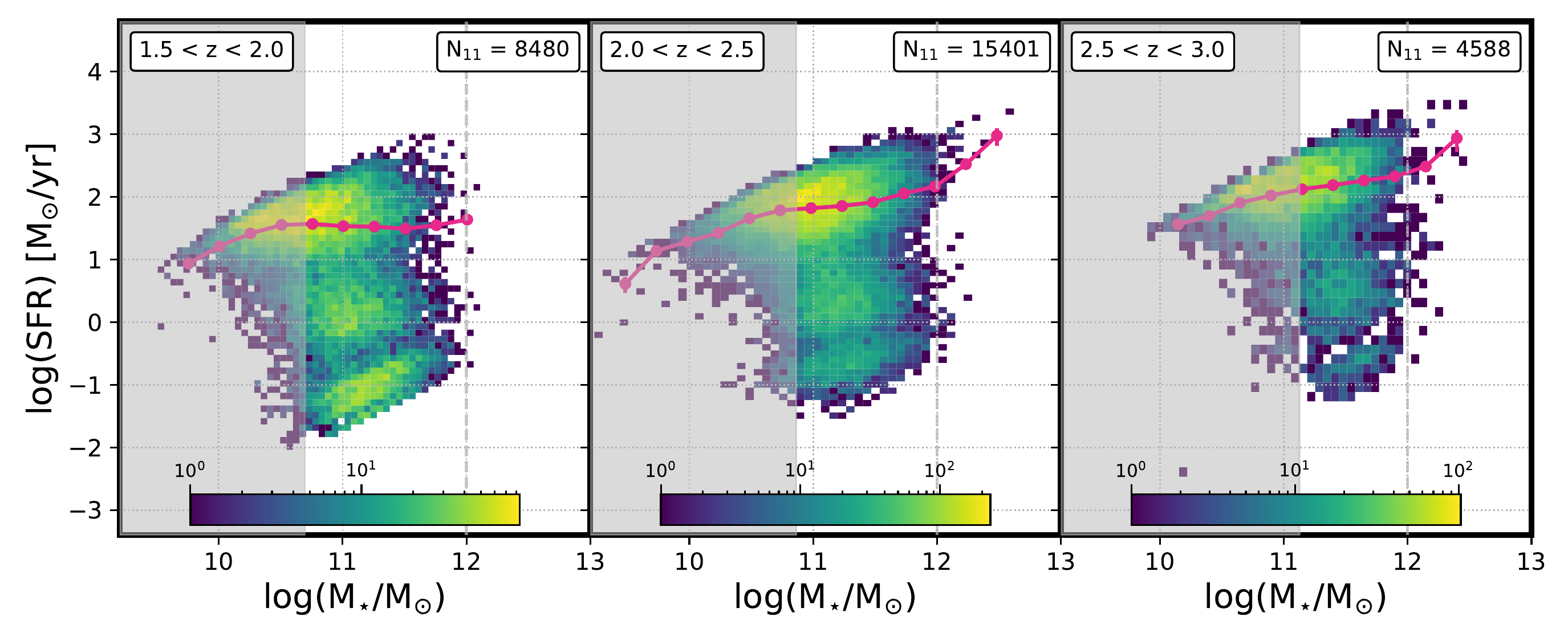} 
\caption{The SFR-M$_{\star}$ relation (2D histogram) and main sequence (pink circles) for all galaxies in our sample. The main sequence is the average SFR in individual mass bins, while errors on the main sequence are computed using the bootstrap resampling procedure described in Section \ref{sec:MainSequence}. The main sequence for all galaxies shows a flattening at the highest masses ($M_\star = 10^{11}$ to $10^{12}$M$_\odot$), and this flattening becomes more prominent as time progresses towards $z=1.5$. Colorbars show the number of galaxies in each cell of the 2D histogram, and gray shaded regions represent masses below our 95\% completeness limit. We emphasize that the results presented in this work focus on the mass range $M_\star = 10^{11}$ to $10^{12}$M$_\odot$, and that results above $M_\star = 10^{12}M_\odot$ (vertical dashed gray line) are unlikely to be robust. Insets on the upper right of each panel show the total number ($N_{11}$) of galaxies in our sample with $M_\star \ge 10^{11}$M$_\odot$.}
\label{ms_all_plot}
\end{center} 
\end{figure*}

We find that at the high mass end ($M_\star = 10^{11}$ to $10^{12}$M$_\odot$), the main sequence for all galaxies is flattened (Figure \ref{ms_all_plot}; compared to the often assumed slope of unity; e.g., \citealt{Wuyts2011}), and this flattening becomes more pronounced as redshift decreases toward $z=1.5$. Although we do not assume any functional form of the main sequence, using an ordinary least squares regression (fit to main sequence values between $M_\star = 10^{11}$ to $10^{12}$M$_\odot$), we can determine that the power law slope of the main sequence evolves from $0.30\pm0.0005$ at $2.5 < z < 3.0$, to $0.24\pm0.0008$ at $2.0 < z < 2.5$, and finally to $-0.02\pm0.0004$ at $1.5 < z < 2.0$. Further exploration of the implication of the shape of the main sequence for all galaxies will be discussed in Sections \ref{sec:MSAll_vs_MSSF} and \ref{sec:Discussion}.

%%%%%%%%%%%%%%%%%%%%%% MS All Galaxies Figure

%%%%%%%%%%%%%%%%%%%%%% MS SF Galaxies Subsection
\subsection{Isolating Star-Forming Galaxies and Measuring the Main Sequence for Star-Forming Galaxies}
\label{sec:SF_MainSequence}
To compute the star-forming galaxy main sequence, we first need to isolate the star-forming galaxy population from the collective green valley and quiescent population (see Figure \ref{ms_all_plot}). To do this, we require a method that both utilizes the quantities of interest in this study (stellar mass and star-formation rate) and does not place artificial limits on the width or scatter around the star-forming galaxy main sequence, as that would limit our ability to study the scatter around this relation later in this work (see Section \ref{sec:MS_scatter}).

\cite{Sherman2020b} used three methods to separate the star-forming and quiescent galaxy populations: a fixed specific star-formation rate (sSFR; $\rm sSFR = \rm SFR/\rm M_\star$) threshold, a fixed distance below the main sequence, and UVJ color-color selection. All three methods give quiescent fractions as a function of mass that are consistent within a factor of two. The fixed sSFR and distance below the main sequence methods both place artificial limits on the scatter around the main sequence by using a fixed threshold separating star-forming and quiescent galaxies. In \cite{Sherman2020b} the sSFR threshold was set to be sSFR = 10$^{-11} \rm yr^{-1}$ for all mass bins in our three redshift bins spanning $1.5 < z < 3.0$. Using the distance from the main sequence method, \cite{Sherman2020b} computed the main sequence in the same way as described here and considered all galaxies lying 1 dex or more below the main sequence to be quiescent. This method was an improvement over the fixed sSFR threshold because the threshold varied with stellar mass and redshift bin, however it still set an artificial limit of 1dex on the scatter around the star-forming galaxy main sequence. Alternatively, the UVJ color-color method seeks to separate galaxies into star-forming and quiescent populations by using their position in color-color parameter space. These populations were initially interpreted using evolutionary tracks (e.g., \citealt{Labbe2005}, \citealt{Wuyts2007}), and a boundary was later placed between them using the empirically-based locations of the two populations (e.g., \citealt{Williams2009}, \citealt{Muzzin2013}). Although this method is a common way to separate star-forming and quiescent galaxies (e.g., \citealt{Whitaker2014}, \citealt{Tomczak2016}) it relies heavily on where the boundary between star-forming and quiescent galaxies is drawn and how rest-frame U, V, and J fluxes are estimated during the SED fitting procedure. 

An unbiased, meaningful way of isolating the star-forming galaxy population would be to employ the information provided by the SFR-M$_{\star}$ plane itself. Our large sample size allows us to make this separation by locating the transition between star-forming galaxies and galaxies in the green valley (e.g., \citealt{Martin2007}, \citealt{Salim2007}, \citealt{Wyder2007}) in individual small mass bins. In this work, we consider green valley galaxies to be those lying in the region of the SFR-M$_{\star}$ plane below the star-forming galaxy population and above the quenched galaxy population. We note that while some works select green valley galaxies in color space, we exclusively refer to this population as it relates to their location in the SFR-M$_{\star}$ plane. 

Previous works with significantly smaller samples than ours have studied the green valley population by separating transitional green valley galaxies from star-forming and quiescent populations in the SFR-M$_{\star}$ plane. \cite{Pandya2017} made this separation at $z=0-3$ by first finding the main sequence (where the normalization is determined using the $M_\star = 10^{9} - 10^{9.5} $M$_\odot$ population and the slope is assumed to be unity), then defining a region from $0.6 -1.4$ dex below the main sequence which contained the green valley population. \cite{Jian2020} first found the median relationships for all star-forming and quiescent galaxies (where the former is simply the star-forming main sequence and the latter is a linear fit to the quiescent galaxy sample, where these populations are found using an iterative approach) and defined the center of the green valley to be the average of these linear fits for a sample of galaxies at $z=0.2-1.1$. They then adopted a fixed width for the green valley to define their transition galaxy population, and the upper limit of this region served as a fixed lower limit for the star-forming population. Both the methods from \cite{Pandya2017} and \cite{Jian2020} place artificial limits on the width of the star-forming galaxy population in the SFR-M$_{\star}$ plane, the same limitation encountered in \cite{Sherman2020b}. 

A different, yet similarly limiting approach, is taken by \cite{Janowiecki2020} who define the star-forming population at $z=0.01-0.05$ by fitting un-constrained Gaussians to the sSFR distributions of galaxies in small mass bins. This is first done at low masses where galaxies are predominantly star-forming, then the modes of these Gaussians are extrapolated to higher masses to define the main sequence around which one-sided Gaussians with fixed modes are then fit to galaxies with sSFR greater than the mode. The star-forming population is defined to be the Gaussian distribution of galaxies around the star-forming main sequence (extrapolated modes), and they define green valley galaxies to be those $1\sigma$ below the ridge of the main sequence. Because the width of the best-fitting Gaussian in a given mass bin is determined solely from fitting a one-sided Gaussian to galaxies lying above the extrapolated mean sSFR, the lower bound of the star-forming population is reliant on the distribution of highly star-forming galaxies and the underlying assumption that star-forming galaxies adhere to a Gaussian distribution in the SFR-M$_{\star}$ plane. 

%%%%%%%%%%%%%%%%%%%%%% GV Top Schematic
\begin{figure}
\begin{center}
\includegraphics[width=3.3in]{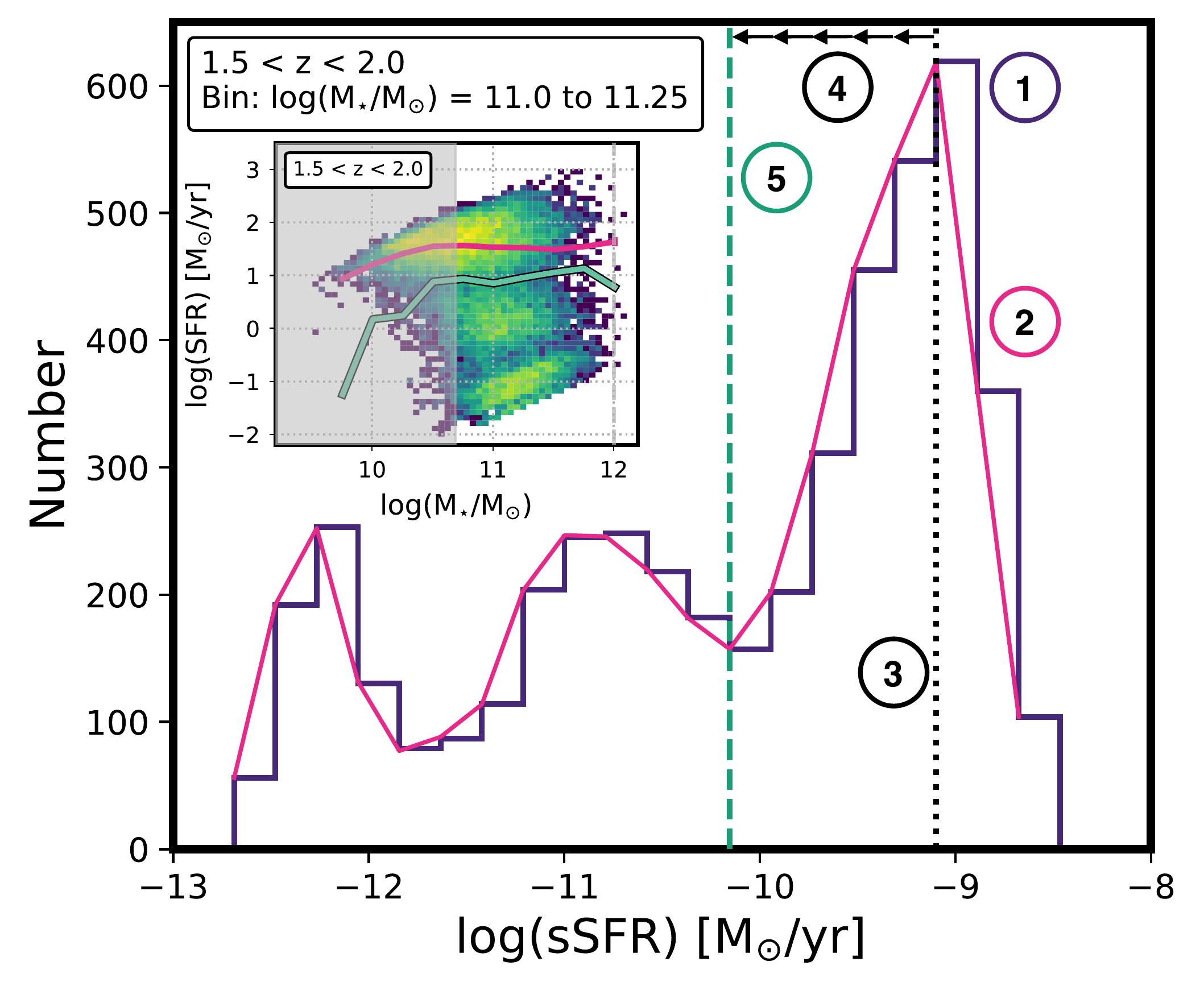} 
\caption{An example schematic of our method used to locate the transition between the star-forming and green valley galaxy populations. The labeled steps are as follows and they correspond to the same numbered steps in Section \ref{sec:SF_MainSequence}. Step 1: For all galaxies in a given mass bin (in this example, the $M_\star = 10^{11}$M$_\odot$ bin for $1.5 < z < 2.0$ galaxies) construct a histogram of specific star-formation rate values. Step 2: Interpolate the shape of this histogram using a univariate spline. Step 3: Find the local maximum at log(sSFR)$>-10.2$ as a rough estimate of the ridge of the main sequence. Step 4: Step bin-by-bin from higher to lower sSFR. Step 5: Stop bin-by-bin stepping when the interpolated spline goes from decreasing to increasing, and define this local minimum as the transition between the star-forming and green valley populations. We remind the reader that for every galaxy in our sample, we obtain a measure of dust-corrected SFR from our SED fitting procedure. The inset figure shows the SFR-$M_\star$ plane in the $1.5 < z < 2.0$ bin with the main sequence for all galaxies shown in pink and the dividing line (green with black outline) between the star-forming and green valley populations determined using the procedure described here and in Section \ref{sec:SF_MainSequence}. The results of implementing this procedure to isolate the star-forming population in all three redshift bins can be seen in Figure \ref{ms_sf_plot}. In the inset figure, the gray shaded region represents masses below our 95\% completeness limit, and the vertical dashed gray line represents $M_\star = 10^{12}M_\odot$, above which our results are unlikely to be robust.}
\label{gv_annotated_plot}
\end{center} 
\end{figure} 

In this work, we avoid biasing the scatter around the star-forming galaxy main sequence and use the values of interest (stellar mass and star-formation rate) to isolate the star-forming galaxy population, by employing a method that locates the transition between the star-forming galaxy population and galaxies lying in the green valley. We locate this transition in each of our small mass bins (mass bins have 0.25 dex width) within our three redshift bins spanning $1.5 < z < 3.0$ in order to isolate the star-forming galaxy population without using fixed cutoffs. We are uniquely suited to take this approach because each of our small mass bins contains enough high mass galaxies to robustly locate the transition between the star-forming and green valley populations.

We note that the local minima seen in the SFR-$M_\star$ plane (see Figure \ref{ms_all_plot}) are physically motivated, and they are not products of our SED fitting procedure. Our SED fitting method determines the best-fitting SED for each galaxy by combining a set of twelve SED templates in non-negative linear combination. There are no constraints placed on the contribution of each template, aside from requiring that the templates either provide a positive contribution or zero contribution to the final best-fitting SED. Therefore, since galaxies are fit to be in the transition regions between populations, these regions of parameter space are accessible to these template combinations. If galaxies are not fit to be in the transition regions between populations it is because those regions did not provide the best-fitting SED, not because those regions are inaccessible to the template SEDs. 

Locating the transition between the star-forming and green valley populations is a five step process (see Figure \ref{gv_annotated_plot} for a schematic). First, in each of our small mass bins we construct a histogram of the sSFR of all galaxies in that bin. These histograms are binned using the Freedman-Diaconis Estimator \citep{Freedman1981}, which optimizes bin size based on sample size while being robust to outliers. This allows the bin size for each sSFR histogram in small mass bins to vary based on the number of galaxies in that small mass bin. Second, we interpolate over this histogram using a univariate spline with degree three (a cubic spline). This smoothed interpolation is robust to small amounts of bin-to-bin noise, and allows us to define the shape of the sSFR histogram in each small stellar mass bin. We note that the spline interpolation is based on the left edges of the sSFR histogram bins because we want all galaxies placed in an sSFR bin to have the same designation as star forming or green valley. If we were to place the separation between the star-forming and green valley populations at an sSFR in the center of an sSFR bin, then the galaxies in that bin would be placed into two separate categories. Third, we estimate the location of the ridge of the main sequence by finding the local maximum at log(sSFR)$>-10.2$, and fourth we step along our interpolated distribution from high to low sSFR values until the number of galaxies switches from decreasing to increasing. This switch occurs at the local minimum between the star-forming population and the green valley population, and finally (step five), we define this local minimum to be the transition between star-forming galaxies and the collective green valley and quiescent galaxy population in each of our small mass bins.

We note that the procedure described above is only implemented when there are more than 100 galaxies in a small mass bin and the transition between the star-forming and green valley populations can be clearly defined. This required number of galaxies was determined through trial and error. We found that when there were fewer than 100 galaxies in a given mass bin, the sSFR histogram was too sparsely populated to reliably locate the transition between star-forming and green valley galaxies, if any exists. In that case, the threshold between star-forming and quiescent galaxies was set to sSFR = 10$^{-11} \rm yr^{-1}$. This only impacts mass bins well below our completeness limit or at the extreme high-mass end ($M_\star > 10^{12}$M$_\odot$), and, because this work focuses on the mass range $M_\star = 10^{11}$ to $10^{12}$M$_\odot$, this requirement of 100 galaxies in a small mass bin does not impact our results. 

A potential source of uncertainty in isolating star-forming galaxies with this method arises from measurement uncertainty. Our method relies on accurately identifying the first inflection point leftward of the main sequence in the sSFR histogram in a given mass bin. If the true sSFR value for a galaxy is slightly different than the sSFR measured from our SED fitting procedure, the true inflection point in the sSFR histogram may be different than the one we measure. To investigate the impact of this type of uncertainty, we implement a procedure in which we draw a new sSFR (and associated stellar mass) for every galaxy in our science catalog using its parameter measurement errors given by our SED fitting procedure (see \citealt{Sherman2020} for details of this error measurement). We then repeat the above procedure to re-compute the location of the inflection point in the sSFR histogram in each mass bin. This procedure is performed 1000 times, thereby giving 1000 values, in each mass bin, of the local minimum between the star-forming and green valley populations. We are then able to investigate how different inflection point locations impact our measurements of the main sequence for star-forming galaxies and the scatter around that relation. Through this procedure we find that the typical draw gives an sSFR inflection point within a factor of $\sim2$ of our best-fit measurement for $M_\star = 10^{11}$ to $10^{12}$M$_\odot$. Because there are relatively few galaxies around the local minimum between the star-forming and green valley populations, we find that our measured main sequence for star-forming galaxies and the scatter around that relation are robust (within factors of $\sim1.3$ and $\sim2$, respectively) to small changes in the value for the local minimum in the sSFR histogram. Although we allow galaxies to move between mass bins during this test, we note that our best-fit measurements of the (star-forming) galaxy main sequence and scatter around the star-forming galaxy main sequence, which are presented throughout this work, do not account for scatter between mass bins (we remind the reader that typical stellar mass errors are $\pm0.08$ dex for $1.5 < z < 3.0$ galaxies above our estimated mass completeness limits, which is significantly smaller than our 0.25 dex bin size). 

\begin{figure}
\begin{center}
\includegraphics[width=3.3in]{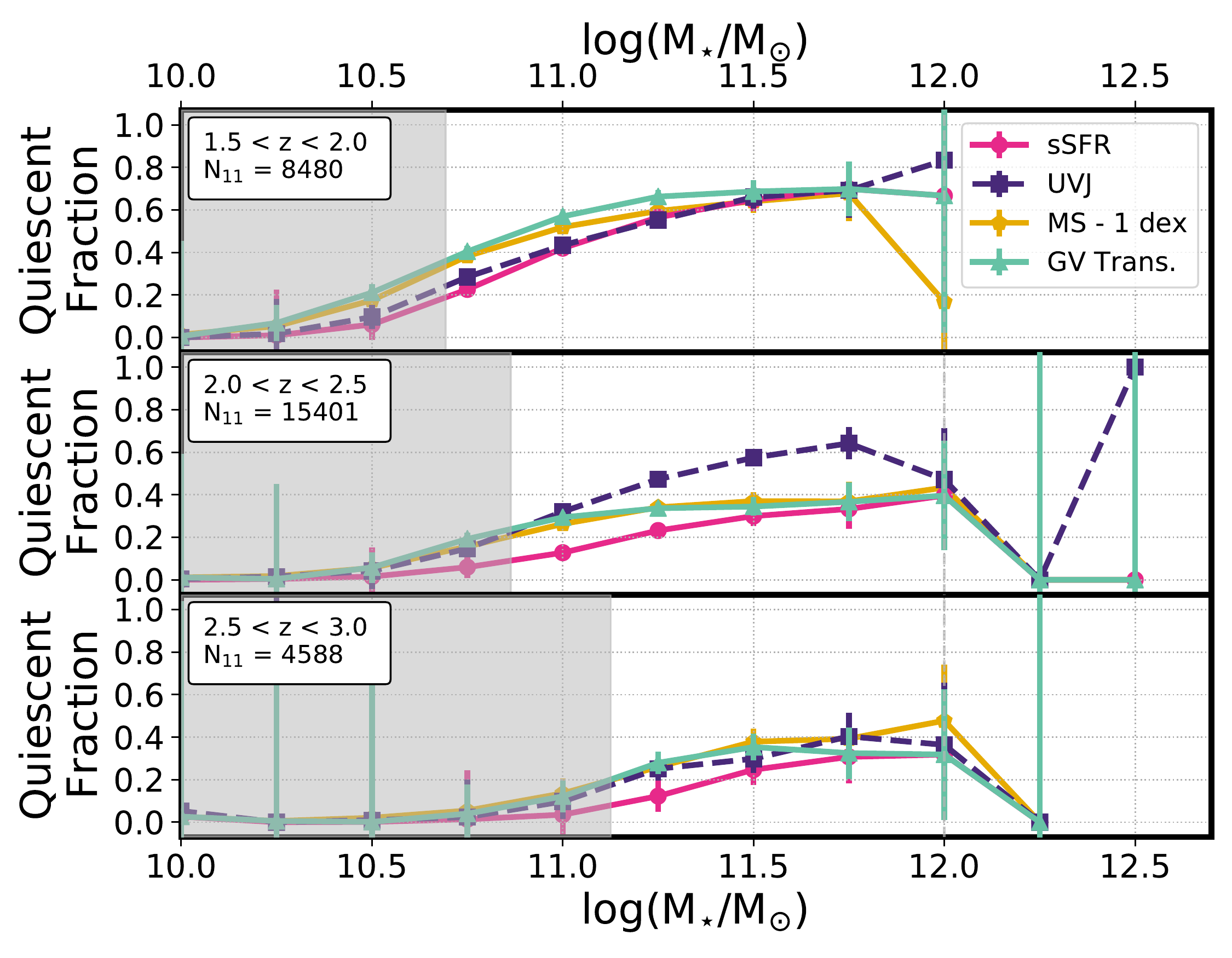} 
\caption{The quiescent fraction as a function of stellar mass determined using the transition between star-forming and green valley galaxies to separate star-forming systems from the collective green valley and quiescent populations (green triangles). Also plotted are the results from \protect\cite{Sherman2020b} who determined the quiescent fraction in three ways: sSFR-selected (pink circles), main sequence - 1 dex selected (gold pentagons), and UVJ-selected (purple squares). The four measurements of the quiescent fraction give consistent results across our three redshift bins spanning $1.5 < z < 3.0$. Gray shaded regions represent masses below our 95\% completeness limit. Error bars represent Poisson errors. We emphasize that the results presented in this work focus on the mass range $M_\star = 10^{11}$ to $10^{12}$M$_\odot$, and that results above $M_\star = 10^{12}M_\odot$ (vertical dashed gray line) are unlikely to be robust. Insets on the upper left of each panel show the total number ($N_{11}$) of galaxies in our sample with $M_\star \ge 10^{11}$M$_\odot$.}
\label{qf_comp}
\end{center} 
\end{figure}

To confirm that our method separating star-forming galaxies from the collective population of green valley and quiescent galaxies is consistent with other methods of isolating star-forming galaxies, we compare our collective fraction of green valley and quiescent galaxies to the fractions determined by \cite{Sherman2020b}, who used three methods (sSFR-selected, main sequence - 1 dex selected, and UVJ-selected quiescent fractions; Figure \ref{qf_comp}). The agreement is strongest with the quiescent fraction computed using the main sequence - 1 dex method. This is expected as this method was most effective at separating star-forming galaxies from the collective green valley and quiescent galaxy populations in \cite{Sherman2020b}. The method implemented in this work is an improvement over the main sequence - 1 dex method as it does not place an arbitrary distance below the main sequence as a criterion for isolating star-forming galaxies.

%%%%%%%%%%%%%%%%%%%%%% Quiescent fraction comparison

\begin{figure*}
\begin{center}
\includegraphics[width=\textwidth]{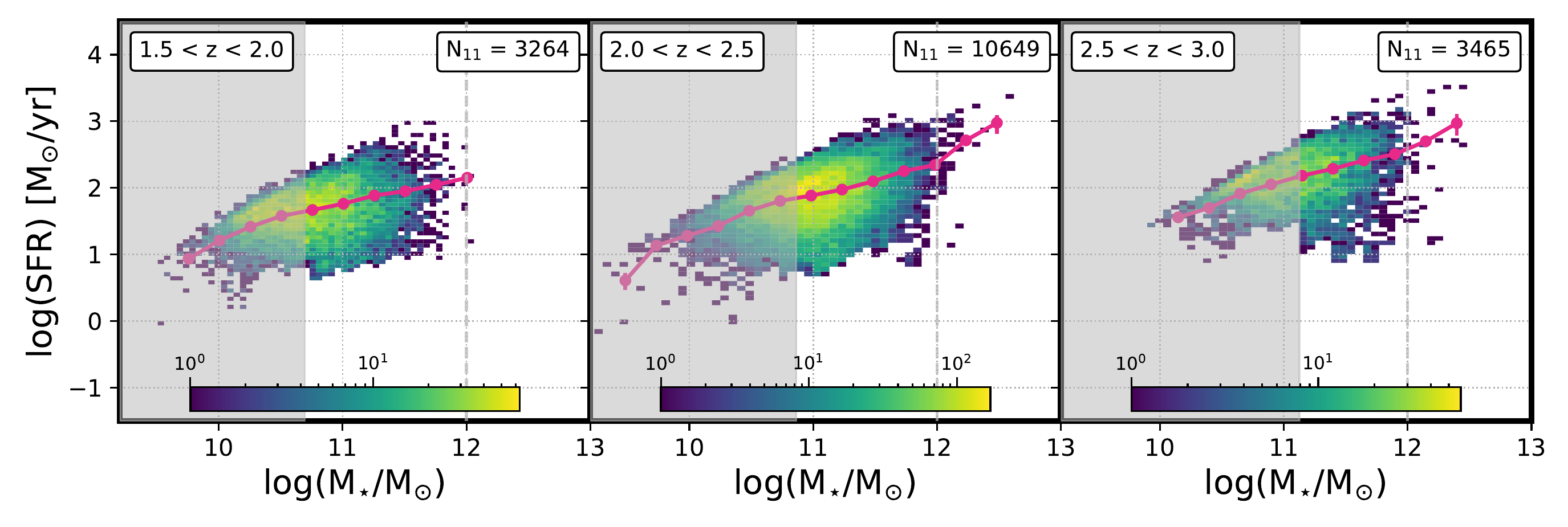} 
\caption{The SFR-M$_{\star}$ relation (2D histogram) and main sequence (pink circles) for star-forming galaxies in our sample. Star-forming galaxies are selected by locating the transition between star-forming and green valley populations, then removing galaxies below this transition, as is described in Section \ref{sec:SF_MainSequence}. The star-forming main sequence is the average SFR in individual mass bins, while errors on the star-forming main sequence are computed using the bootstrap resampling procedure described in Section \ref{sec:MainSequence}. Unlike the main sequence for all galaxies, the star-forming galaxy main sequence does not show a strong evolution in the high mass end slope from $z=3.0$ to $z=1.5$. Colorbars show the number of galaxies in each cell of the 2D histogram, and gray shaded regions represent masses below our 95\% completeness limit. We emphasize that the results presented in this work focus on the mass range $M_\star = 10^{11}$ to $10^{12}$M$_\odot$, and that results above $M_\star = 10^{12}M_\odot$ (vertical dashed gray line) are unlikely to be robust. Insets on the upper right of each panel show the number ($N_{11}$) of star-forming galaxies in our sample with $M_\star \ge 10^{11}$M$_\odot$.}
\label{ms_sf_plot}
\end{center} 
\end{figure*}

With a population of star-forming galaxies identified, we are able to compute the star-forming galaxy main sequence (Figure \ref{ms_sf_plot}), which is the average SFR in each mass bin, with error bars computed using the bootstrap resampling procedure described in Section \ref{sec:MainSequence}. The star-forming galaxy main sequence does not show a significant flattening at the high mass end ($M_\star = 10^{11}$ to $10^{12}$M$_\odot$). Its power law slope, computed using an ordinary least squares fit to the star-forming galaxy main sequence values over the mass range $M_\star = 10^{11}$ to $10^{12}$M$_\odot$, evolves mildly from $0.47\pm0.0011$ at $2.5 < z < 3.0$, to $0.46\pm0.0001$ at $2.0 < z < 2.5$ , and finally to $0.35\pm0.0013$ at $1.5 < z < 2.0$.

%%%%%%%%%%%%%%%%%%%%%% MS Star-Forming Galaxies Figure

\subsection{Implications of the Growing Green Valley and Quiescent Populations}
\label{sec:MSAll_vs_MSSF}
\begin{figure*}
\begin{center}
\includegraphics[width=6in]{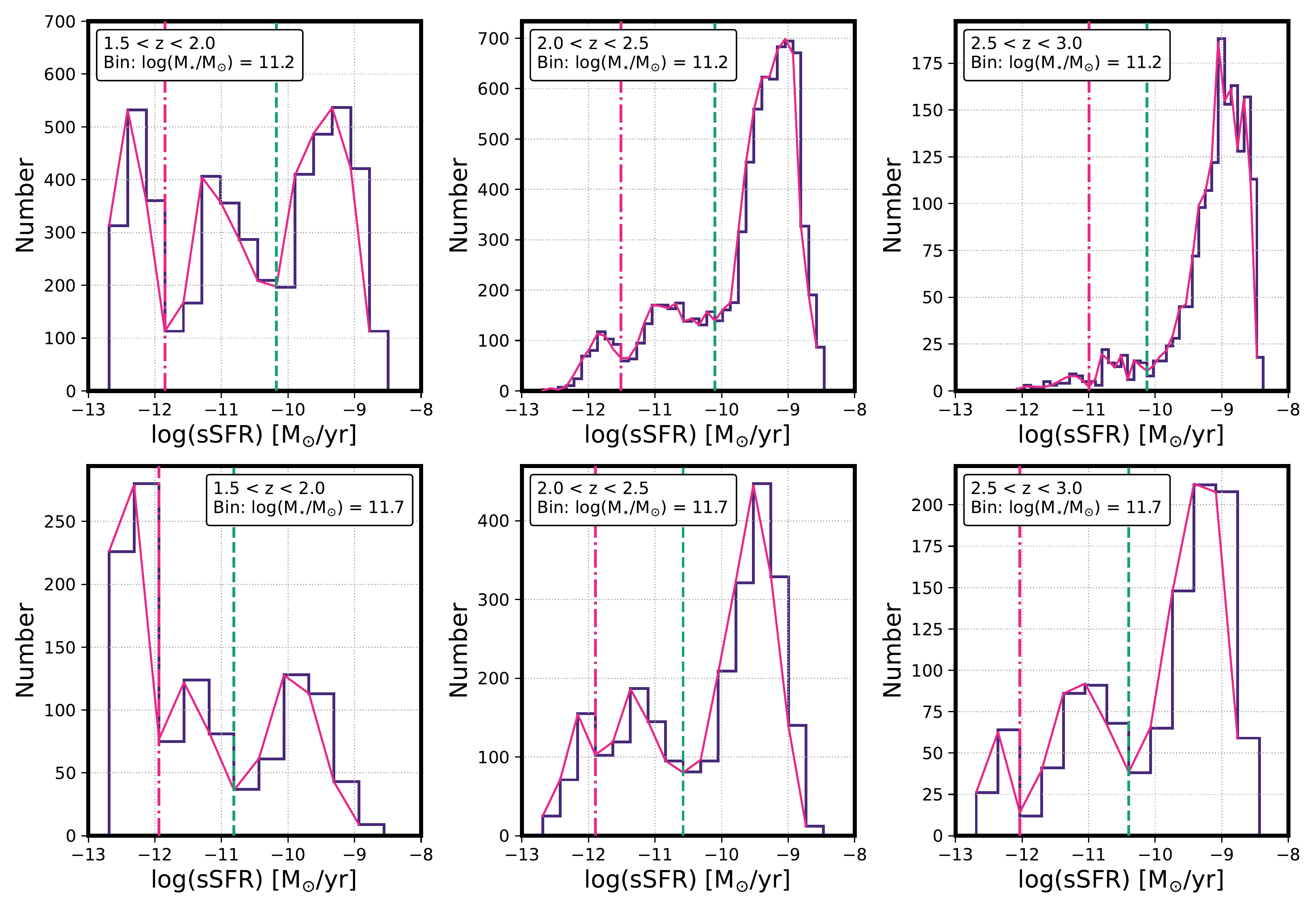} 
\caption{Specific star-formation rate distributions for individual mass bins in the SFR-$M_\star$ plane (purple histograms), with the splines used to interpolate these distributions (solid pink lines). The three vertical columns of panels are for each of our three redshift bins spanning $z=1.5$ to $z=3.0$. The top row shows the log($M_{\star}$/$\rm M_{\odot}$) = 11.2 bin, and the bottom row shows the log($M_{\star}$/$\rm M_{\odot}$) = 11.7 bin. In each panel, star-forming galaxies fall to the right of the vertical dashed green line, green valley galaxies are between the vertical dashed green line and the vertical dash-dot pink line, and quiescent galaxies lie to the left of the vertical dash-dot pink line. The procedure used to identify the location of the transition between star-forming and green valley galaxies and transition between green valley and quiescent galaxies are described in Sections \ref{sec:SF_MainSequence} and \ref{sec:MSAll_vs_MSSF}, respectively. As we move from higher to lower redshift, the buildup of the populations of green valley and quiescent galaxies becomes prominent. We again note that our SED fitting procedure provides a measure of dust-corrected SFR for every galaxy in our $K_s$-selected sample. }
\label{gvTrans_moreBins_plot}
\end{center} 
\end{figure*}

As is seen in Figure \ref{ms_all_plot}, our large sample of galaxies in the SFR-$M_\star$ plane shows three distinct populations of galaxies: star-forming, green valley, and quiescent. In Section \ref{sec:SF_MainSequence} we described a novel method for using the transitions between these populations to isolate the star-forming galaxy population. This procedure can also be used to find the local minimum in sSFR space between the green valley and quiescent populations. To locate this transition, we employ a version of the five step procedure described in Section \ref{sec:SF_MainSequence}, with a small modification to step three. Here, we (1) construct an sSFR histogram in each mass bin, (2) interpolate using a smoothed cubic spline, (3) find the local maximum of the green valley population (local maximum between log(sSFR)$>-12.0$ and the sSFR at which the local minimum occurs between the green valley and star-forming populations, as determined in Section \ref{sec:SF_MainSequence}), (4) step bin-by bin from high to low sSFR, and finally (5) stop stepping when a local minimum in the spline is found. This local minimum is the transition between the green valley and quiescent populations (Figure \ref{gvTrans_moreBins_plot}).

In Figure \ref{gvTrans_moreBins_plot}, we show that the sSFR distributions in individual mass bins can provide more information about the buildup of the collective green valley and quiescent populations as time progresses and that higher mass bins have larger collective populations of quiescent and green valley galaxies than star-forming galaxies. This result is consistent with measures of the quiescent fraction from \cite{Sherman2020b}, who showed that at these redshifts and stellar masses, the quiescent fraction increases from $z=3.0$ to $z=1.5$ at the highest masses and that higher mass galaxies ($M_\star = 10^{12}$M$_\odot$) at a given redshift have a larger quiescent fraction than lower mass systems ($M_\star = 10^{11}$M$_\odot$). The method used in this work to isolate star-forming galaxies by locating the transition between star-forming and green valley galaxies is an improvement over the main sequence - 1 dex technique used by \cite{Sherman2020b} as it more meaningfully isolates star-forming galaxies from the collective green valley and quiescent population without employing an ad hoc threshold below the main sequence. 

%%%%%%%%%%%%%%%%%%%%%% sSFR histograms

\begin{figure}
\begin{center}
\includegraphics[width=3.3in]{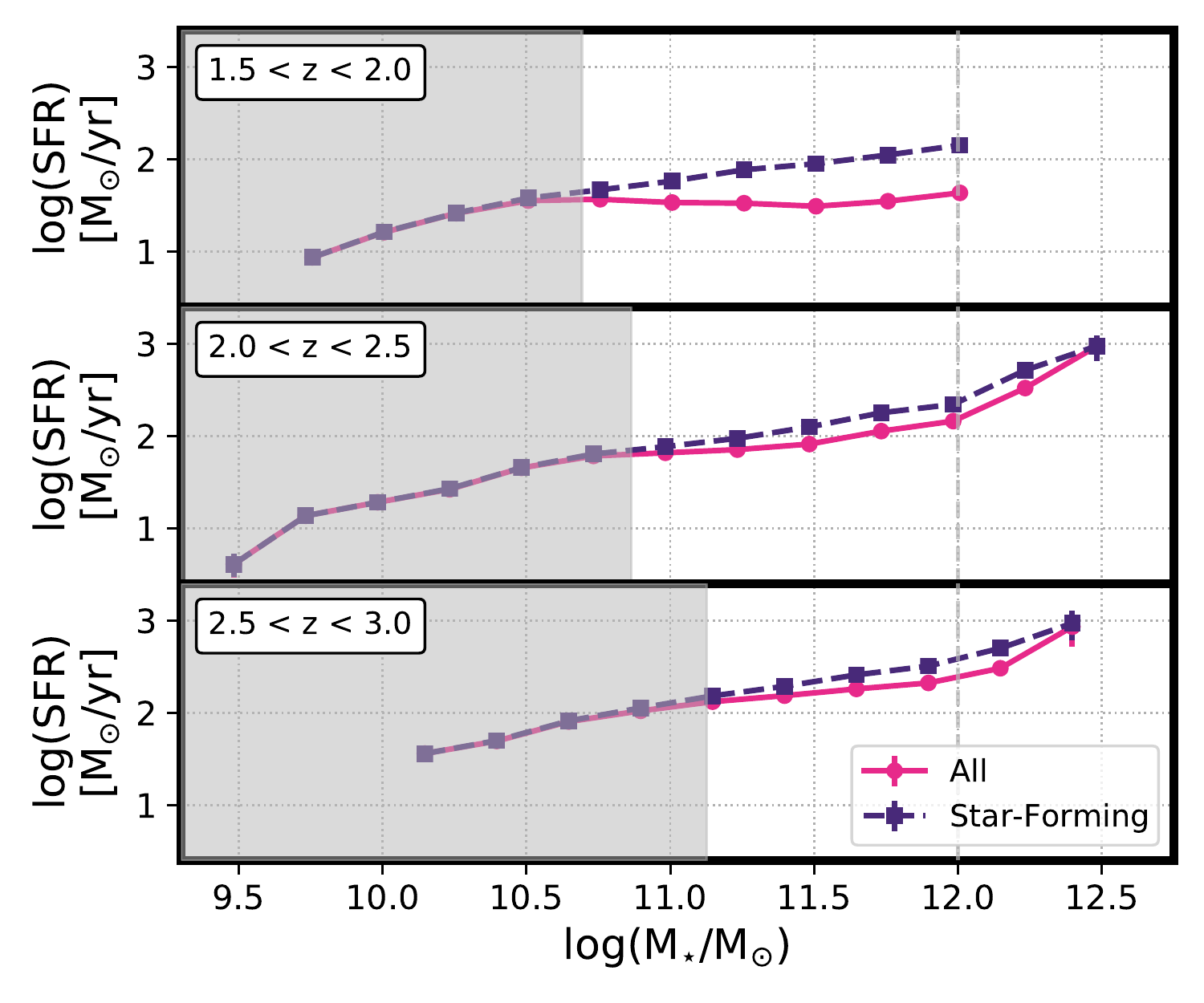} 
\caption{The main sequence for all galaxies (pink circles) and star-forming galaxies (purple squares) in our sample. Star-forming galaxies are selected by locating the transition between star-forming and green valley populations, then removing galaxies below this transition, as is described in Section \ref{sec:SF_MainSequence}. The (star-forming) main sequence is the average SFR in individual mass bins, while errors on the (star-forming) main sequence are computed using the bootstrap resampling procedure described in Section \ref{sec:MainSequence}. We note that error bars are included, however they are often smaller than the symbol. At early epochs ($z > 2$) the star-forming galaxy main sequence is up to a factor of 1.5 higher than the main sequence for all galaxies, and at later epochs ($1.5 < z < 2.0$), the star-forming galaxy main sequence is a factor of $1.5 - 3$ higher than the main sequence for all galaxies. Gray shaded regions represent masses below our 95\% completeness limit. We emphasize that the results presented in this work focus on the mass range $M_\star = 10^{11}$ to $10^{12}$M$_\odot$, and that results above $M_\star = 10^{12}M_\odot$ (vertical dashed gray line) are unlikely to be robust.}
\label{ms_all_v_sf_plot}
\end{center} 
\end{figure}

Our empirical main sequences measured for all galaxies (see Section \ref{sec:MainSequence}) and star-forming galaxies (see Section \ref{sec:SF_MainSequence}) are compared in Figure \ref{ms_all_v_sf_plot}. In our two highest redshift bins ($2.0 < z < 2.5$ and $2.5 < z < 3.0$), where only $\sim20-40\%$ of massive ($M_\star \ge 10^{11}$M$_\odot$) galaxies are members of the collective green valley and quiescent population (Figure \ref{qf_comp}), the total galaxy main sequence is higher than the star-forming galaxy main sequence by up to a factor of 1.5. At lower redshifts ($1.5 < z < 2.0$) where the collective green valley and quiescent population are $\sim40-70\%$ of the total massive galaxy population (Figure \ref{qf_comp}), the star-forming galaxy main sequence is a factor of $1.5-3$ higher than the main sequence for the total galaxy population. The significant buildup of the collective green valley and quiescent galaxy populations as a function of redshift and stellar mass leads to the flattening of the massive end slope of the main sequence for all galaxies as time progresses from $z=3.0$ to $z=1.5$.

%%%%%%%%%%%%%%%%%%%%%% MS All vs Star-Forming

Our sample, which is used to study both the main sequence for all galaxies and star-forming galaxies contains galaxies with $M_\star > 10^{12}$M$_\odot$, particularly in the $2.0 < z < 2.5$ and $2.5 < z < 3.0$ bins where the comoving volume observed by our study is larger. For this extreme high-mass population, we see main sequence relations with steeper slopes at $M_\star > 10^{12}$M$_\odot$ than are seen at stellar masses $M_\star = 10^{11}$ to $10^{12}$M$_\odot$. Individual mass bins above $M_\star = 10^{12}$M$_\odot$ have fewer than 100 galaxies, making robust studies of this population challenging. \cite{Sherman2020b} also showed that the impact of uncertainties in photometric redshifts and Eddington bias on results for this extreme population is likely to be large (see \citealt{Sherman2020b} and their Appendix Figure A1) and that some of this population may be low-redshift interlopers. In this work, we focus on galaxies in the mass range $M_\star = 10^{11}$ to $10^{12}$M$_\odot$, and note that high-resolution imaging and spectroscopic followup of these extreme high-mass objects is necessary to better understand their properties and behavior in the SFR-$M_\star$ plane.  

%%%%%%%%%%%%%%%%%%%%%%%%%%%%%%%%%%%%%%%%%%%%%%%%%%
%%%%%%%%%%%%%%%%% MS Scatter %%%%%%%%%%%%%%%%%%
%%%%%%%%%%%%%%%%%%%%%%%%%%%%%%%%%%%%%%%%%%%%%%%%%%
\section{Scatter Around the Star-Forming Galaxy Main Sequence} 
\label{sec:MS_scatter}
In the absence of stochastic processes (e.g., mergers, gas accretion from the cosmic web, stellar and AGN feedback), the relationship between stellar mass and star-formation rate for star-forming galaxies should be relatively tight, with scatter around that relationship due only to measurement uncertainty (e.g., \citealt{Caplar2019},  \citealt{Matthee2019}). Therefore, measures of the scatter around the star-forming galaxy main sequence provide insights into the importance of stochastic processes in driving galaxy evolution. \cite{Sherman2020b} outlined how different stochastic processes could play a key role in driving the evolution of the massive galaxy population, where mergers are likely drivers of early mass buildup and environmental processes (e.g., ram pressure stripping, tidal stripping, harassment) are likely to suppress star-formation at $z < 2$, when emerging clusters develop their intracluster medium (ICM).

\begin{figure*}
\begin{center}
\includegraphics[width=\textwidth]{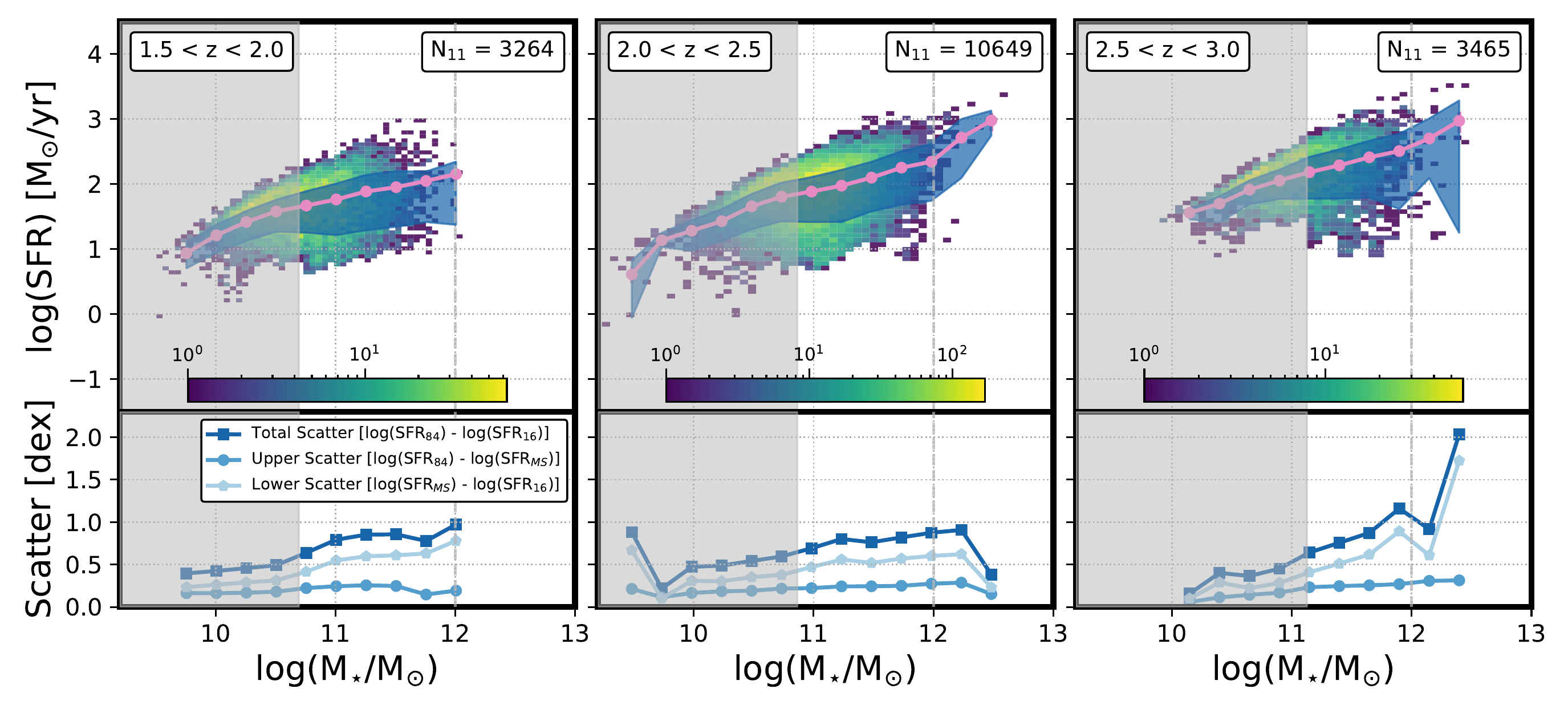} 
\caption{Top Row: The total scatter (blue shaded region) around the star-forming galaxy main sequence (pink) overlaid on the distribution of star-forming galaxies (2D histogram; colorbar indicates the number of galaxies in each 2D bin) in the SFR-$M_\star$ plane. The upper (lower) bound of the blue shaded region is the 84th (16th) percentile of the SFR distribution in a given mass bin. Insets on the upper right of each panel in the top row show the number ($N_{11}$) of galaxies in the star-forming population in our sample with $M_\star \ge 10^{11}$M$_\odot$. Bottom Row: The total (squares), upper (circles), and lower (pentagons) observed scatter around the star-forming galaxy main sequence. The total scatter shows a modest increase with increasing stellar mass (less than a factor of three from $M_\star = 10^{11}$ to $10^{12}$M$_\odot$ in each redshift bin), and the total scatter is fairly constant across our three redshift bins from $z=1.5$ to $z=3.0$. In every redshift bin, the lower scatter is larger than the upper scatter by up to a factor of 3. Gray shaded regions represent masses below our 95\% completeness limit. We emphasize that the results presented in this work focus on the mass range $M_\star = 10^{11}$ to $10^{12}$M$_\odot$, and that results above $M_\star = 10^{12}M_\odot$ (vertical dashed gray line) are unlikely to be robust. }
\label{ms_scatter}
\end{center} 
\end{figure*}

We measure the total scatter around the star-forming galaxy main sequence (Figure \ref{ms_scatter}) without assuming either a functional form of the main sequence or a fixed criterion for isolating star-forming galaxies. This is a significant improvement over previous studies where the selection of the star-forming galaxy population was biased and measures of the scatter often assumed an underlying distribution of galaxies in the SFR-$M_\star$ plane (such as a Gaussian; see Section \ref{subsec:Scatter_CompWObs} for further comparison with previous empirical results). As is described in Section \ref{sec:SF_MainSequence}, our star-forming galaxy population is selected by locating the transition between the star-forming and green valley populations in small mass bins, and the star-forming galaxy main sequence is the average SFR of the star-forming galaxy population in each mass bin. This approach is made possible by our large sample of 28,469 massive ($M_\star \ge 10^{11}$M$_\odot$) galaxies spanning $1.5 < z < 3.0$. 

The total scatter around the star-forming galaxy main sequence measured in each of our small mass bins is simply the difference between the 84$^{\rm th}$ and 16$^{\rm th}$ percentile of the distribution of SFR values for star-forming galaxies in each mass bin. We also compute the upper scatter (difference between 84$^{\rm th}$ percentile of SFR and the star-forming galaxy main sequence value in a given mass bin) and lower scatter (difference between the star-forming galaxy main sequence value and 16$^{\rm th}$ percentile of SFR in a given mass bin) to provide a closer comparison with previous works. 

Additionally, we can approximate the intrinsic scatter around the star-forming galaxy main sequence by accounting for the $\pm0.18$ dex measurement uncertainty in SFR from our SED fitting procedure. Our SFR error estimates are determined by drawing 100 SEDs from the best-fit SED's template error distribution (see \cite{Sherman2020} for a detailed description of this procedure), and therefore, this error estimate takes into account uncertainties in other fundamental measurements, such as extinction. We do not find that the measurement uncertainty in SFR varies as a function of stellar mass, indicating that removing the scatter due to measurement uncertainty will not change the trends (or lack thereof) observed in the total, upper, and lower scatter as a function of mass and redshift.  

We measure the total observed scatter to be $\sim0.5-1.0$ dex (corresponding to $\sim0.47-0.98$ dex intrinsic scatter) and we find that the total observed scatter increases from low to high masses ($M_\star = 10^{11}$ to $10^{12}$M$_\odot$) by less than a factor of three in each of our three redshift bins. The scatter does not show significant evolution as a function of redshift across our three redshift bins spanning $1.5 < z < 3.0$. 

In each of our redshift bins, the observed upward scatter is fairly constant as a function of mass and redshift, with a value of $\sim0.3$ dex (corresponding to $\sim0.24$ dex intrinsic scatter), consistent with values for the observed scatter found by previous studies (see Section \ref{subsec:Scatter_CompWObs}). The lower scatter around the main sequence is larger than the upper scatter in all redshift bins. Our result shows that the often assumed symmetrical Gaussian distribution of star-forming galaxies around the star-forming galaxy main sequence does not hold true at these redshifts ($1.5 < z < 3.0$) for massive ($M_\star \ge 10^{11}$M$_\odot$) galaxies. 

%%%%%%%%%%%%%%%%%%%%%% MS Scatter

%%%%%%%%%%%%%%%%%%%%%%%%%%%%%%%%%%%%%%%%%%%%%%%%%%
%%%%%%%%%%%%%%%%% Comparison with Observations %%%%%%%%%%%%%%%%%%
%%%%%%%%%%%%%%%%%%%%%%%%%%%%%%%%%%%%%%%%%%%%%%%%%%
\section{Comparison With Previous Observations} 
\label{sec:CompWObs}

%%%%%%%%%%%%%%%%%%%%%% MS Comp w/ Obs Subsection
\subsection{Comparison of the Main Sequence for All Galaxies and Star-Forming Galaxies with Previous Observations}
\label{subsec:MS_CompWObs}
Since the first work referencing the galaxy main sequence by \cite{Noeske2007}, many works have implemented different methods of measuring the main sequence for all galaxies and isolating star-forming galaxies to measure the star-forming galaxy main sequence. Rapid innovation in galaxy surveys over the past decade has produced a number of new methods, however this makes true one-to-one comparisons with previous works difficult. 

In this work we have leveraged our sample of massive galaxies, the largest uniformly selected set compiled to date, to measure the galaxy main sequence for the total population and star-forming galaxies in small mass bins at the high mass end. A significant benefit of our large sample is that we do not need to adopt a functional form of the main sequence, and we can isolate star-forming galaxies in a meaningful way without prior assumptions. Moreover, the large area probed by our study also renders errors due to cosmic variance negligible. 

\begin{figure*}
\begin{center}
\includegraphics[width=\textwidth]{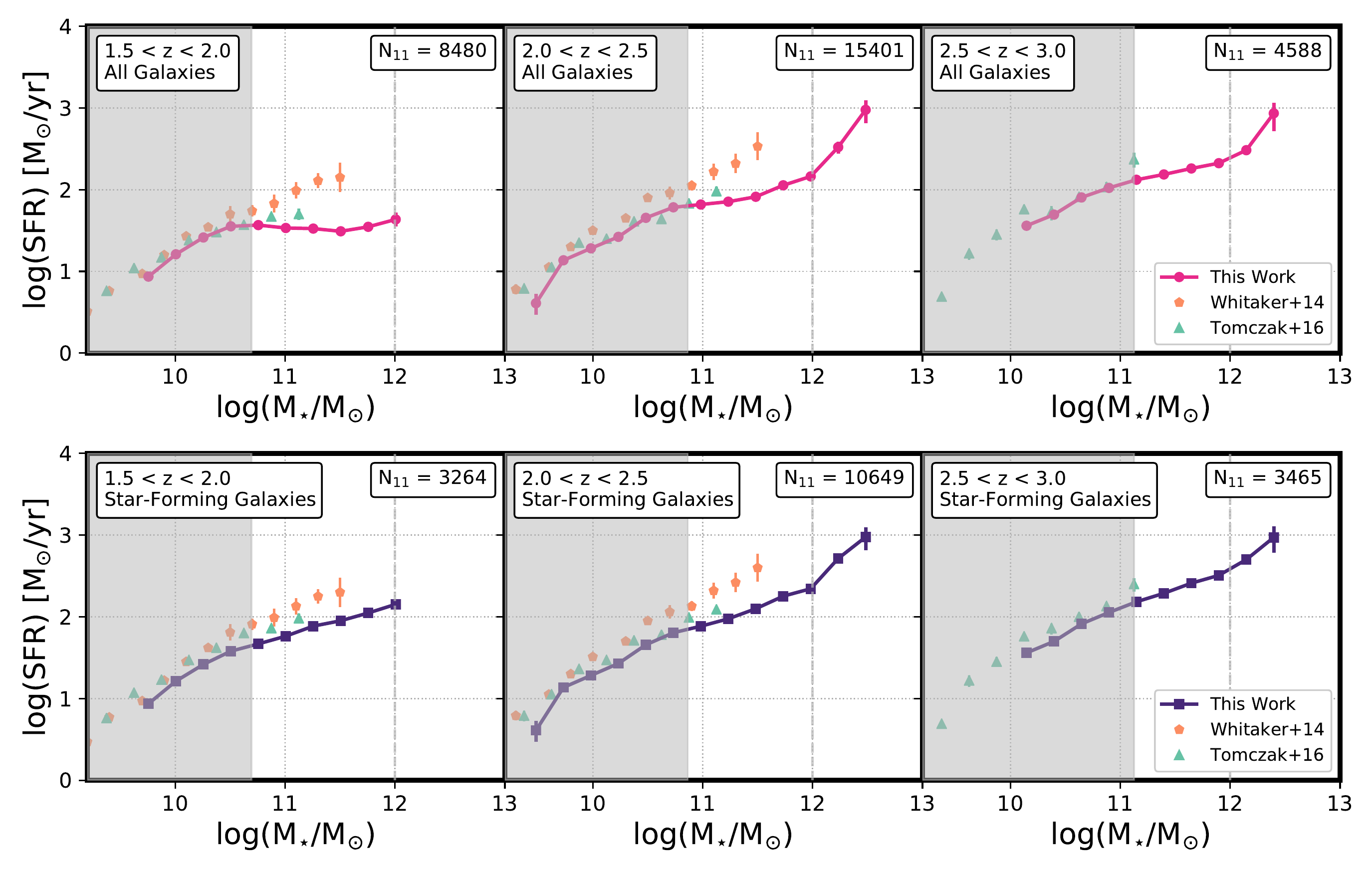} 
\caption{Our empirical main sequence for all galaxies (top row) and star-forming galaxies (bottom row) compared with results from previous observations. We find similar results to those from \protect\cite{Tomczak2016} for both the total (top row) and star-forming (bottom row) galaxy populations. The results from \protect\cite{Whitaker2014} for both the total (top row) and star-forming (bottom row) galaxy populations are higher than our empirical results by a factor of $\sim1.5-6.5$. Gray shaded regions represent masses below our 95\% completeness limit. Insets on the upper right of each panel show the number ($N_{11}$) of galaxies for the total population (top row) and star-forming population (bottom row) in our sample with $M_\star \ge 10^{11}$M$_\odot$. We emphasize that the results presented in this work focus on the mass range $M_\star = 10^{11}$ to $10^{12}$M$_\odot$, and that results above $M_\star = 10^{12}M_\odot$ (vertical dashed gray line) are unlikely to be robust.}
\label{ms_vObs_plot}
\end{center} 
\end{figure*}

The two works we compare with (\citealt{Whitaker2014} and \citealt{Tomczak2016}) present the main sequence for both the total galaxy population and star-forming galaxy population. The study from \cite{Whitaker2014} focused on the low-mass end of the main sequence using galaxies in the CANDELS/3D-HST fields. Their main sequence values in individual mass bins were computed using stacked UV + IR luminosities (with $L_{\rm IR}$ from \textit{Spitzer}-MIPS 24$\mu m$ photometry and assuming a \cite{Chabrier2003} IMF), and they separated star-forming and quiescent galaxies using UVJ colors. Their work finds that the main sequence is best characterized by a broken power law fit, however for comparison with our empirical result (Figure \ref{ms_vObs_plot}), we utilize their average stacked SFR values in each stellar mass bin rather than the functional fit to those data. We find that the main sequence for all galaxies and star-forming galaxies from \cite{Whitaker2014} are factors of $1.5-4.5$ and $1.7-3$ higher than our empirical main sequences for all galaxies and star-forming galaxies, respectively at $1.5 < z < 2.5$ for $M_\star = 10^{11}$ to $10^{12}$M$_\odot$. The study from \cite{Whitaker2014} does not investigate the main sequence in our highest redshift bin ($2.5 < z < 3.0$). 

\cite{Tomczak2016} performed a similar study to that from \cite{Whitaker2014}, using a stacking analysis of UV + IR luminosities (also with $L_{\rm IR}$ from \textit{Spitzer}-MIPS 24$\mu m$ photometry and using a \cite{Chabrier2003} IMF) to derive the average SFR (main sequence values) in small mass bins for galaxies in ZFOURGE. Similar to \cite{Whitaker2014}, \cite{Tomczak2016} separated star-forming and quiescent galaxies using UVJ color. When comparing with our empirical result, we find that the results from \cite{Tomczak2016} are in general agreement, within a factor of $\sim1.5$, with our main sequence for all galaxies and star-forming galaxies in our three redshift bins spanning $1.5 < z < 3.0$ for $M_\star = 10^{11}$ to $10^{12}$M$_\odot$. We note that in the $2.5 < z < 3.0$ bin, the highest masses probed by \cite{Tomczak2016} only reach our mass completeness limit. Therefore, comparisons between our empirical result and that from \cite{Tomczak2016} in the $2.5 < z < 3.0$ bin are not informative. 

There are several important caveats to the comparisons presented above that must be noted. First, these studies both utilize data from similar legacy fields, often the same fields with updated photometry, spectroscopy, or different modeling techniques. The CANDELS/3D-HST fields ($\sim900$ arcmin$^2$) used by \cite{Whitaker2014} include AEGIS, COSMOS, GOODS-N, GOODS-S, and UDS. The ZFOURGE fields ($\sim400$ arcmin$^2$) used by \cite{Tomczak2016} include CDF-S, COSMOS, and UDS. Second, these legacy fields, while rich in spectroscopy and multi-wavelength photometry allowing for strongly constrained SEDs, are small area studies with small samples of galaxies at the highest masses. Across the three redshift bins spanning $1.5 < z < 3.0$, the study from \cite{Tomczak2016} has 81 $M_\star \ge 10^{11}$M$_\odot$ galaxies in their total galaxy population. The publicly available CANDELS/3D-HST catalog (\citealt{Brammer2012}, \citealt{Skelton2014}) used by \cite{Whitaker2014} has 533 $M_\star \ge 10^{11}$M$_\odot$ total galaxies spanning $1.5 < z < 2.5$, however \cite{Whitaker2014} may have only used a sub sample of these objects. Third, the small areas probed by these studies may be strongly impacted by the effects of cosmic variance. For $M_\star \ge 10^{11}$M$_\odot$ the cosmic variance is $\sim50-70\%$ for studies of this size \citep{Moster2011}. For comparison, our 17.5 deg$^2$ study has 28,469 $M_\star \ge 10^{11}$M$_\odot$ galaxies between $1.5 < z < 3.0$. This effectively eliminates errors due to cosmic variance. Finally, the stacked 24$\mu$m-based SFR values used by \cite{Whitaker2014} and \cite{Tomczak2016} may be systematically different from our rest-frame UV-based SFR values determined for individual galaxies through SED fitting. 

%%%%%%%%%%%%%%%%%%%%%% MS vs Obs

%%%%%%%%%%%%%%%%%%%%%% Scatter Comp w/ Obs Subsection
\subsection{Comparison of the Scatter Around the Star-Forming Galaxy Main Sequence with Previous Observations} 
\label{subsec:Scatter_CompWObs}
Using our method of isolating the star-forming galaxy population by locating the transition between the star-forming and green valley populations in the SFR-$M_\star$ plane, we investigated the scatter around the star-forming galaxy main sequence in Section \ref{sec:MS_scatter}. Our result shows that the scatter around the star-forming galaxy main sequence does not evolve significantly either as a function of stellar mass or redshift over the stellar mass range $M_\star = 10^{11}$ to $10^{12}$M$_\odot$ and redshifts $1.5 < z < 3.0$. Our method of isolating star-forming galaxies does not place artificial constraints on the lower boundary of this population, or on the underlying distribution of star-forming galaxies in the SFR-$M_\star$ plane, and we find that star-forming galaxies are not normally distributed around the star-forming galaxy main sequence. This is a significant finding as the distribution of star-forming galaxies in the SFR-$M_\star$ plane is often assumed to be a Gaussian by previous works. 

Comparisons with previous results for the scatter around the star-forming galaxy main sequence are challenging as a consensus has not been reached by previous works when it comes to measuring the scatter. These measurements are further complicated by the different approaches to separating star-forming galaxies from the total population (e.g. different color indicators or fixed thresholds) and different ways of measuring the main sequence (e.g., stacking analyses, average SFR, median SFR, extrapolation from low to high masses, assumed functional forms). 

\cite{Schreiber2015} investigated the scatter of galaxies above the main sequence using individual \textit{Herschel}-detected galaxies in the CANDELS-\textit{Herschel} fields out to $z=4$. They found the scatter above the main sequence to be 0.32 dex with little evolution as a function of mass or redshift. \cite{Rodighiero2011} used a sample of $1.5 < z < 2.5$ BzK color-selected star-forming galaxies in COSMOS and found the scatter around the main sequence to be 0.24 dex, assuming a Gaussian distribution of galaxies around the star-forming galaxy main sequence. \cite{Popesso2019} took yet another approach, whereby they used an IR-selected sample of star-forming galaxies in the CANDELS+GOODS fields out to $z=2.5$ and only fit for the normalization of the main sequence, adopting the slope from the local relation. They found that the scatter around the star-forming galaxy main sequence increases from $\sim0.3$ dex to $\sim0.4$ dex as a function of mass for $1.5 < z < 2.5$ galaxies. 

Results from these previous studies are broadly consistent with our result, however detailed comparisons are difficult due to the very different methods used. Additionally, as described above, our approach to measuring the scatter around the star-forming galaxy main sequence is a significant improvement over previous works as it does not rely on assumed functional forms of the star-forming galaxy main sequence, ad hoc cutoffs for selecting the star-forming galaxy population, or assuming a Gaussian distribution of star-forming galaxies in the SFR-$M_\star$ plane. 

%%%%%%%%%%%%%%%%%%%%%%%%%%%%%%%%%%%%%%%%%%%%%%%%%%
%%%%%%%%%%%%%%%%% Comparison with Theory %%%%%%%%%%%%%%%%%%
%%%%%%%%%%%%%%%%%%%%%%%%%%%%%%%%%%%%%%%%%%%%%%%%%%
\section{Comparison With Theoretical Models} 
\label{sec:CompWTheory}
In this work, we have explored the main sequence for all galaxies (Section \ref{sec:MainSequence}), used a novel approach to identify the star-forming galaxy population in the SFR-M$_\star$ plane (Section \ref{sec:SF_MainSequence}), and investigated the scatter around the star-forming galaxy main sequence (Section \ref{sec:MS_scatter}) in an unbiased way, with our focus placed on the mass range $M_\star = 10^{11}$ to $10^{12}$M$_\odot$. Theoretical models, such as hydrodynamical simulations and semi-analytic models (SAMs), seek to implement physical processes that drive galaxy evolution and, therefore, insights from theoretical models may allow for interpretation of the physical processes driving observed trends. Large volume empirical studies, such as the study presented in this work, can likewise provide benchmarks for these models.  

Our comparison will focus on the hydrodynamical models SIMBA \citep{Dave2019} and IllustrisTNG (\citealt{Pillepich2018b}, \citealt{Springel2018}, \citealt{Nelson2018}, \citealt{Naiman2018}, \citealt{Marinacci2018}), as well as the semi-analytic model SAG \citep{Cora2018}. Details about each model can be found in their respective publications, as well as in \cite{Sherman2020b}, and the key points will briefly be described here. SIMBA has a 100 Mpc/h box with mass resolution $m_{\rm gas} = 1.82\times10^7~M_{\odot}$ and we utilize the total stellar mass and SFR for each galaxy in their group catalog. IllustrisTNG offers several volumes, and we use the largest box that is $\sim$300$^3$Mpc$^3$ (TNG300) with mass resolution $m_{\rm baryon}=1.1\times10^7~M_{\odot}$ and masses and SFR measured within twice the stellar half mass radius (the $2 \times R_{1/2}$ aperture; see \citealt{Sherman2020b} for a detailed study of aperture types in IllustrisTNG). SAG populates halos in the MultiDark-Planck2 (MDPL2) dark matter-only simulation, and we utilize an updated version of the model (S. Cora, private communication) which has been run on 9.4\% of the 1.0 $h^{-1}$Gpc box available in MDPL2. For SAG, we use total masses and SFR for galaxies in their group catalog. The group catalogs for all three models hard code SFR = 0 when the SFR for an object falls below the resolution limit. To account for this, following \cite{Donnari2019} and \cite{Sherman2020b}, we assign these objects a random SFR between SFR = $10^{-5} - 10^{-4}$M$_{\odot} yr^{-1}$ before performing our analysis. 

The above models have significantly smaller volumes than our empirical study, leading to significantly smaller numbers of galaxies with stellar masses $M_\star = 10^{11}$ to $10^{12}$M$_\odot$. Because of this, we will compare our empirical main sequence for all galaxies with the corresponding relation from the theoretical models, with a focus on galaxies with masses spanning the range $M_\star = 10^{11}$ to $10^{12}$M$_\odot$. At this time, a fair and informative comparison cannot be done between our empirical star-forming galaxy main sequence and results from theoretical models, as the theoretical models do not have enough galaxies in small mass bins spanning $M_\star = 10^{11}$ to $10^{12}$M$_\odot$ across our redshift range of interest ($1.5 < z < 3.0$) to define the relation or study the scatter around it. 

The main sequence for all galaxies for each of the three theoretical models is computed in the same way as our empirical main sequence (see Section \ref{sec:MainSequence}) We find that in all three of our redshift bins spanning $1.5 < z < 3.0$, the hydrodynamical model SIMBA is in fair agreement, within a factor of $\sim1.5$, with our empirical main sequence for all galaxies, but does not show a flattening at the highest masses in the lowest redshift bin ($1.5 < z < 2.0$). The SAM SAG is up to a factor of $\sim3$ higher than our empirical result and starts to show a flattening slope at the highest masses in the $1.5 <  z < 2.0$ bin. The hydrodynamical model IllustrisTNG is below our empirical result by up to a factor of $\sim10$ and shows a strong turnover at the highest masses in the $2.0 < z < 3.0$ bins, where our empirical result does not show this trend. These results are consistent with those from \cite{Sherman2020b} who showed that SIMBA and SAG under-estimate the fraction of the collective green valley and quiescent galaxy population compared with the star-forming galaxy population at the high-mass end, while the IllustrisTNG model was shown to over-predict the fraction of massive galaxies lying in the collective green valley and quiescent galaxy population. 

Further exploration of the implication of comparisons with results from theoretical models will be discussed in Section \ref{sec:Discussion}. We also recognize that a single line (in this case, the main sequence for all galaxies) is not an adequate representation of the full distribution of galaxies in the SFR-$M_\star$ plane, and we refer the reader to Appendix \ref{app:TheoryComp} for a comparison of our empirical data in the SFR-$M_\star$ plane and those from the models.

%%%%%%%%%%%%%%%%%%%%%% MS vs Theory
\begin{figure*}
\begin{center}
\includegraphics[width=\textwidth]{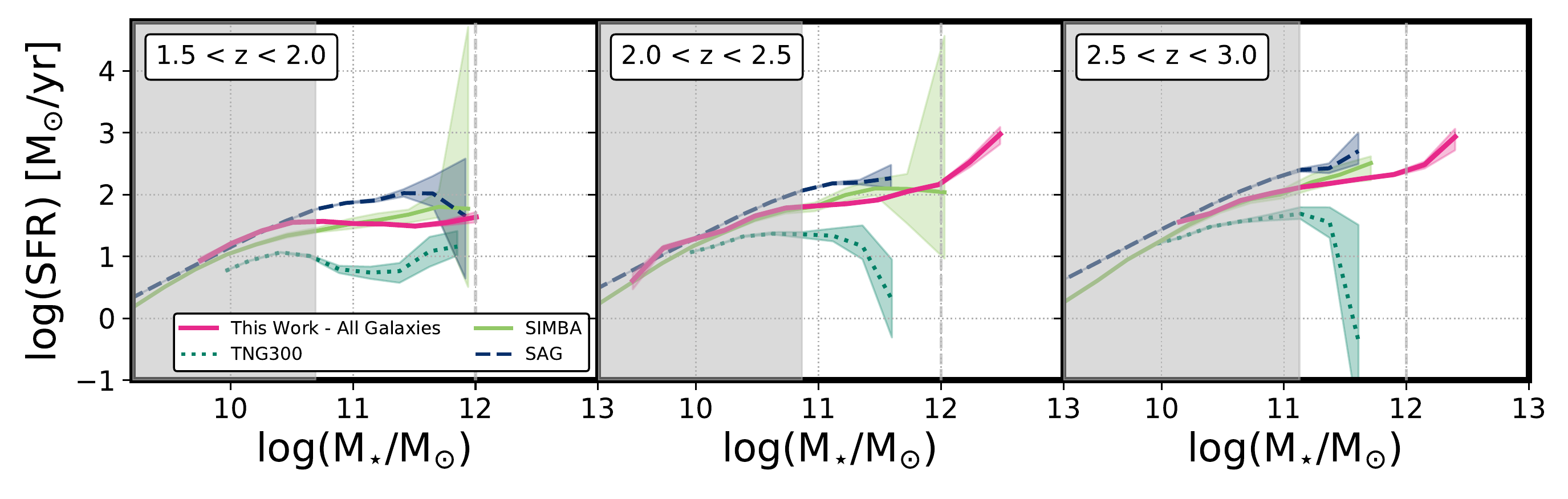} 
\caption{Our empirical main sequence for all galaxies compared with results from hydrodynamical models SIMBA and IllustrisTNG and SAM SAG. The main sequence for all galaxies from SIMBA is within a factor of $\sim1.5$ of our empirical result and that from SAG is higher than our empirical result by up to a factor of $\sim3$. SIMBA does not show a flattening at the highest masses by $z=1.5$, while SAG begins to show a flattening high-mass slope towards $z=1.5$. The main sequence for all galaxies from IllustrisTNG is lower than our empirical result by up to a factor of $\sim10$ and shows a strong turnover at the highest masses at $2.0 < z < 3.0$ that is not seen in our empirical result. Gray shaded regions represent masses below our 95\% completeness limit. We emphasize that the results presented in this work focus on the mass range $M_\star = 10^{11}$ to $10^{12}$M$_\odot$, and that results above $M_\star = 10^{12}M_\odot$ (vertical dashed gray line) are unlikely to be robust.}
\label{ms_vTheory_plot}
\end{center} 
\end{figure*} 

%%%%%%%%%%%%%%%%%%%%%%%%%%%%%%%%%%%%%%%%%%%%%%%%%%
%%%%%%%%%%%%%%%%% Discussion %%%%%%%%%%%%%%%%%%
%%%%%%%%%%%%%%%%%%%%%%%%%%%%%%%%%%%%%%%%%%%%%%%%%%
\section{Discussion} 
\label{sec:Discussion}
In this work, we presented the main sequence for all galaxies (Section \ref{sec:MainSequence}), the main sequence for star-forming galaxies (Section \ref{sec:SF_MainSequence}), and an unbiased measurement of the scatter around the star-forming galaxy main sequence (Section \ref{sec:MS_scatter}). 
Our large sample of massive ($M_\star \ge 10^{11}$M$_\odot$) galaxies has allowed us to separate star-forming galaxies from the green valley and quiescent populations in a natural way without any prior assumptions about the data. With our approach, we are able, for the first time, to present the star-forming galaxy main sequence and measure the scatter about the mean relation without making assumptions about the functional form of the main sequence or the distribution of star-forming galaxies in the SFR-$M_\star$ plane. 

The slope of the main sequence for all galaxies at the high-mass end provides important information about the downsizing \citep{Cowie1996} of the massive galaxy population. \cite{Sherman2020b} showed that the massive ($M_\star \ge 10^{11}$M$_\odot$) galaxy population becomes increasingly quiescent from $z=3.0$ to $z=1.5$, and that more massive galaxies ($M_\star \sim 10^{12}$M$_\odot$) have a higher quiescent fraction than less massive ($M_\star \sim 10^{11}$M$_\odot$) systems. The increased flattening of the high-mass end slope of the main sequence for all galaxies as time progresses from $z=3.0$ to $z=1.5$ traces the downsizing of the massive galaxy population (Figures \ref{ms_all_plot} and \ref{ms_all_v_sf_plot}). A further investigation of the buildup of the collective green valley and quiescent galaxy populations as a function of stellar mass and redshift can be seen in Figure \ref{gvTrans_moreBins_plot}, further supporting the downsizing scenario. 

In contrast to the main sequence for all galaxies, the massive end of the star-forming galaxy main sequence does not demonstrate a strong flattening as time progresses (Figure \ref{ms_all_v_sf_plot}). We have measured that the slope of the massive star-forming galaxy main sequence is rather constant across $1.5 < z < 3.0$, with the slope (and normalization) only beginning to decrease at $1.5 < z < 2.0$. This suggests that, although there is a decrease in the fraction of massive galaxies that are star-forming, those that remain highly star-forming at $1.5 < z < 2.0$ have fairly similar specific star-formation rates as massive star-forming galaxies at earlier epochs ($2.0 < z < 3.0$). 

Our finding that the total scatter around the star-forming galaxy main sequence remains relatively constant from $z=3.0$ to $z=1.5$ and as a function of stellar mass at the high-mass end ($M_\star = 10^{11}$ to $10^{12}$M$_\odot$) is in alignment with our result showing that the slope of the star-forming galaxy main sequence does not significantly flatten as time progresses towards $z=1.5$. The total scatter around the star-forming galaxy main sequence is thought to trace the stochasticity of processes driving star-forming galaxy evolution (e.g., \citealt{Caplar2019},  \citealt{Matthee2019}). Additionally, with our unbiased approach to identifying the star-forming galaxy population, we find that the distribution of massive star-forming galaxies in the SFR-$M_\star$ plane does not follow the often assumed Gaussian distribution. This non-Gaussian distribution around the star-forming galaxy main sequence, which is skewed towards galaxies with lower SFR, suggests that galaxies spend less time in the high SFR phase (above the star-forming galaxy main sequence) than they do in the more moderate SFR phase (below the main sequence). This aligns with studies of the molecular gas content of massive galaxies out to $z\sim4$ using ALMA (e.g., \citealt{Tacconi2018}, \citealt{Franco2020}), which showed that massive galaxies lying above the main sequence have shorter gas depletion timescales than those lying below the main sequence. They report that galaxies above the main sequence may deplete their gas supply in $\sim10^2$ Myr, while star-forming galaxies below the main sequence have depletion times closer to $\sim10^3$ Myr. 

A natural inquiry following the results presented in this work is the question of why the local minima appear in the SFR-$M_\star$ plane between the three populations of interest (star-forming, green valley, and quiescent), and specifically why a peak appears in the green valley. Potential scenarios leading to this green valley peak are complex. Previous works (e.g., \citealt{Pandya2017}, \citealt{Janowiecki2020}) have shown that galaxies do not necessarily take a simple one way trip from the star-forming sequence to the quiescent population, and the way in which galaxies move through the  SFR-$M_\star$ plane, and specifically how they arrive in the green valley population, is dependent on several factors such as environment, available cold gas reservoir, and evolutionary history, among others. This is particularly true for massive galaxies which are likely to live in rich environments where environmental effects are common. Just as environmental mechanisms can remove galaxies from the star-forming population (e.g., AGN and stellar feedback, hot-mode accretion, ram-pressure stripping, tidal stripping, harassment; e.g., \citealt{Man2018}, \citealt{Sherman2020b}), events such as mergers and gas accretion can rejuvenate a previously quenched (or partially quenched) galaxy. The peak seen in the green valley region of the SFR-$M_\star$ plane may arise from the superposition of these massive galaxy populations with diverse evolutionary histories. 

Additionally, the peak in the green valley suggests that galaxies may spend a non-trivial amount of time in this regime. \cite{Pandya2017} estimate that the upper limit for the time galaxies spend in the green valley is $\sim1.5-2$ Gyr for $1.5 < z < 3$. We note, however, that their model makes the simplifying, and unlikely, assumption that galaxies move uni-directionally from the star-forming to quiescent population through the green valley. Interestingly, they find that the population of galaxies in the green valley is rather stable, with more galaxies remaining in the population than moving in or out of the population between timesteps. Again, we emphasize that movement of massive galaxies through and within the green valley is complex and likely to be multi-directional. 

Although our comparisons with theoretical models (hydrodynamical models SIMBA and IllustrisTNG and SAM SAG) are limited to the main sequence for all galaxies due to the small volumes of these models (Section \ref{sec:CompWTheory}), we can use comparisons with these models to interpret the trends seen in our empirical result. The shape and slope of the main sequence for all galaxies, computed for the three theoretical models in the same way as is done for our empirical main sequence for all galaxies, provides information about the transition of the massive galaxy population from being predominantly star-forming to predominantly quiescent. We find in our comparison that the models are unable to simultaneously recover the average specific star-formation rates of massive galaxies found in our observed sample and the flattening of the high-mass end slope of the main sequence for all galaxies as time progresses from $z=3.0$ to $z=1.5$. This result indicates that the models do not adequately represent the observed trends, specifically the buildup of the collective green valley and quiescent populations, seen in our empirical results at these redshifts. Our findings support those from \cite{Sherman2020b} who showed that the SAG and SIMBA models under-estimate the fraction of massive galaxies in the collective green valley and quiescent population, while the IllustrisTNG model over-estimates the fraction of massive galaxies in the collective green valley and quiescent population.

%%%%%%%%%%%%%%%%%%%%%%%%%%%%%%%%%%%%%%%%%%%%%%%%%%
%%%%%%%%%%%%%%%%% SUMMARY %%%%%%%%%%%%%%%%%%
%%%%%%%%%%%%%%%%%%%%%%%%%%%%%%%%%%%%%%%%%%%%%%%%%%
\section{Summary} 
\label{sec:Summary}
Using a large sample of 28,469 massive ($M_\star \ge 10^{11}$M$_\odot$) galaxies that are uniformly selected from data spanning 17.5 deg$^2$, we investigate the nature of the main sequence for all galaxies and for star-forming galaxies. With our large sample, we are uniquely suited to conduct this study without assuming the functional shape of the main sequence or placing prior constraints on the distribution of galaxies in the SFR-M$_{\star}$ plane. A summary of our key results is presented below. 

\begin{enumerate}
\item Our large sample size allows us to compute the main sequence in small stellar mass bins and isolate star-forming galaxies using the quantities of interest (SFR and stellar mass) by finding the local minimum between the star-forming and green valley populations in each mass bin (Fig. \ref{gv_annotated_plot}). A key advantage of this method is that it does not place artificial constraints on the distribution of galaxies around the star-forming galaxy main sequence. Following this approach, we show that the main sequence for all galaxies (Fig. \ref{ms_all_plot}) has a distinct flattening at the high-mass end, which becomes increasingly flat as time progresses from $z=3.0$ to $z=1.5$. We show that this flattening is due to the increasing fraction of the green valley and quiescent galaxy population from $z=3.0$ to $z=1.5$ (Fig. \ref{gvTrans_moreBins_plot}). The star-forming galaxy main sequence (Fig. \ref{ms_sf_plot}) does not show this flattening (see also Fig. \ref{ms_all_v_sf_plot}). This indicates that the average specific star-formation rate of the massive star-forming galaxy population does not evolve significantly over that epoch.
\\
\item We measure the total scatter around the star-forming galaxy main sequence to be $\sim0.5-1.0$ dex and find that there is little evolution in the scatter as a function of stellar mass or redshift (Fig. \ref{ms_scatter}). With our meaningful isolation of star-forming galaxies, we avoid biasing our result by assuming an underlying distribution around the star-forming galaxy main sequence. We also quantify the scatter above (upper) and below (lower) the star-forming galaxy main sequence and find the lower scatter is larger than the upper scatter by up to a factor of 3 in all three redshift bins spanning $1.5 < z < 3.0$, indicating that the underlying distribution of galaxies around the star-forming galaxy main sequence is not the often assumed symmetrical Gaussian.  
\\
\item Additionally, we compare our empirical main sequence for all galaxies with results from theoretical models SIMBA, IllustrisTNG, and SAG. Results from SIMBA are within a factor of $\sim1.5$ of our empirical result but do not show a flattening at the highest masses in our lowest redshift bin ($1.5 < z < 2.0$); those from SAG lie above our results but do show the flattening at the high-mass end in the $1.5 < z < 2.0$ bin. The main sequence for all galaxies from IllustrisTNG is below our empirical result and shows a strong turnover at the highest masses at $2.0 < z < 3.0$, which is not seen in our empirical result. Interpretation of comparisons with theoretical models is not straightforward as the physical processes driving stellar mass buildup and star-formation rates are highly inter-dependent in the models. 
\end{enumerate}

%%%%%%%%%%%%%%%%%%%%%%%%%%%%%%%%%%%%%%%%%%%%%%%%%%
%%%%%%%%%%%%%%%%% ACKNOWLEDGEMENTS %%%%%%%%%%%%%%%%%%
%%%%%%%%%%%%%%%%%%%%%%%%%%%%%%%%%%%%%%%%%%%%%%%%%%
\vspace{5mm}
SS, SJ, and JF gratefully acknowledge support from the University of Texas at Austin, as well as NSF grant AST 1413652.  SS, SJ, JF, and SLF acknowledge support from NSF grant AST 1614798. SS, SJ, JF, MLS, and SLF acknowledge generous support from The University of Texas at Austin McDonald Observatory and Department of Astronomy Board of Visitors. SS, SJ, JF, MLS, and SLF also acknowledge the Texas Advanced Computing Center (TACC) at The University of Texas at Austin for providing HPC resources that have contributed to the research results reported within this paper. SS is supported by the University of Texas at Austin Dissertation Writing Fellowship. MLS and SLF acknowledge support from the NASA Astrophysics and Data Analysis Program through grants NNX16AN46G and 80NSSC18K09. LK and CP acknowledge support from the National Science Foundation through grant AST 1614668. The Institute for Gravitation and the Cosmos is supported by the Eberly College of Science and the Office of the Senior Vice President for Research at the Pennsylvania State University. This publication uses data generated via the Zooniverse.org platform, development of which is funded by generous support, including a Global Impact Award from Google, and by a grant from the Alfred P. Sloan Foundation.

\section*{Data Availability}
No new data were generated in support of this research. We refer the reader to the publications of individual catalogs used in this work for information regarding data access.

%\nocite{*}
\bibliographystyle{mnras}
%\bibliography{references}

%%%%%%%%%%%%%%%%%%%%%%%%%%%%%%%%%%%%%%%%%%%%%%%%%%
%%%%%%%%%%%%%%%%%%%% APPENDIX %%%%%%%%%%%%%%%%%%
%%%%%%%%%%%%%%%%%%%%%%%%%%%%%%%%%%%%%%%%%%%%%%%%%%
\appendix

%%%%%%%%%%%%%%%%%%%%%% Theory SFR Mstar Plane
\renewcommand{\thefigure}{A\arabic{figure}}
\begin{figure*}
\begin{center}
\includegraphics[width=5.8in]{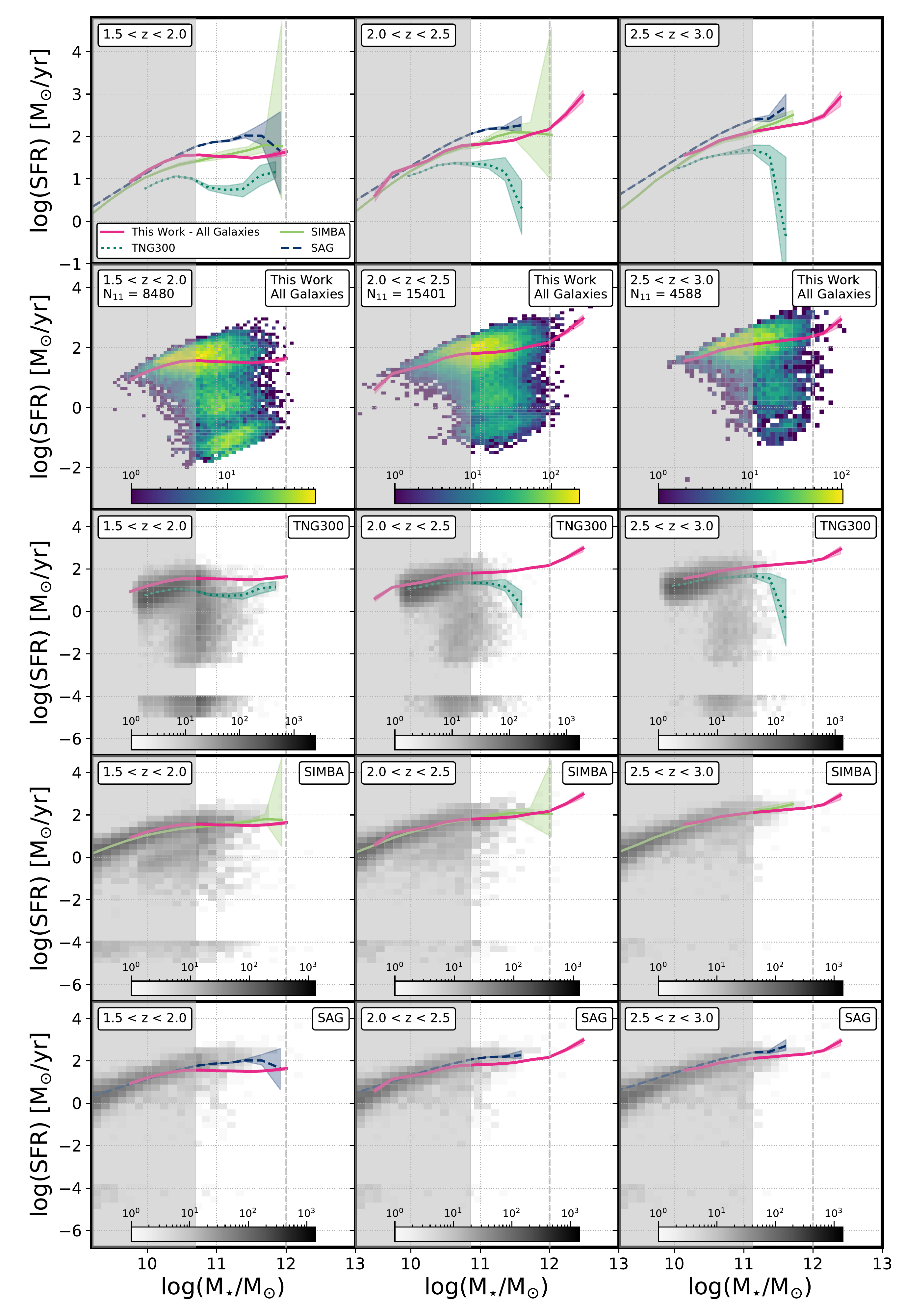} 
\caption{Top row: A reproduction of Figure \ref{ms_vTheory_plot} for ease of comparison. Second row: A reproduction of Figure \ref{ms_all_plot} for ease of comparison. Third to fifth row: The distribution of galaxies in the SFR-$M_\star$ plane for theoretical models IllustrisTNG, SIMBA, and SAG. In rows two through five, the inset box in the upper right corner identifies the data used to generate the 2D histogram in that panel and colorbars show the number of galaxies in each cell of the 2D histogram. IllustrisTNG is more successful at reproducing the distribution of massive galaxies in the SFR-$M_\star$ plane seen in our empirical results than SIMBA and SAG, however IllustrisTNG over-predicts the fraction of galaxies in the collective green valley and quiescent populations \protect\citep{Sherman2020b}. In all rows, the gray shaded regions represent masses below our 95\% completeness limit. We emphasize that the results presented in this work focus on the mass range $M_\star = 10^{11}$ to $10^{12}$M$_\odot$, and that results above $M_\star = 10^{12}M_\odot$ (vertical dashed gray line) are unlikely to be robust.}
\label{sfr_mstar_plane_comp_theory}
\end{center} 
\end{figure*} 

%%%%%%%%%%%%%%%%%%%% TNG APERTURE TEST %%%%%%%%%%%%%%%%%%
\section{The Distribution of Galaxies in the SFR-$M_\star$ Plane for Theoretical Models}
\label{app:TheoryComp} 
In the comparison of our empirical main sequence for all galaxies with those from theoretical models (see Section \ref{sec:CompWTheory}), we noted that the comparison of the main sequence relation alone does not provide all available information about discrepancies (or agreements) between observations and theoretical models. In this Appendix we show (Figure \ref{sfr_mstar_plane_comp_theory}) the full distribution of galaxies in the SFR-$M_\star$ for our data (reproducing Figure \ref{ms_all_plot} for ease of comparison) and the same distributions for the three theoretical models with which we perform comparisons (IllustrisTNG, SIMBA, and SAG). 

We remind the reader that for all three theoretical models, we compute their main sequence from their group catalog data with the same method used for our empirical data (see Section \ref{sec:MainSequence}). The group catalogs for all three models hard code SFR = 0 when the SFR for an object falls below the resolution limit. To account for this, following \cite{Donnari2019} and \cite{Sherman2020b}, we assign these objects a random SFR between SFR = $10^{-5} - 10^{-4}$M$_{\odot} yr^{-1}$ before performing our analysis.

IllustrisTNG is the most successful at producing galaxies in the green valley and quiescent regions of parameter space, however \cite{Sherman2020b} showed that IllustrisTNG over-produces the collective fraction of green valley and quiescent galaxies at $1.5 < z < 3.0$. This translates directly to the main sequence for all galaxies from the IllustrisTNG model having a lower normalization than that from the empirical data and the strong turnover seen at the highest masses in the $2.0 < z < 2.5$ and $2.5 < z < 3.0$ bins. Both SIMBA and SAG were shown by \cite{Sherman2020b} to under-predict the collective fraction of green valley and quiescent galaxies and this can be seen in Figure \ref{sfr_mstar_plane_comp_theory}. 

We emphasize that the ``success'' of a model cannot be tied to a model matching a singular empirical relation well. Instead, several key relations (e.g., the stellar mass function (see \citealt{Sherman2020b}), main sequence, and quiescent fraction (see \citealt{Sherman2020b})) must simultaneously be recovered. The models must also consider the implications that turning individual ``dials'' in the models may have on other measured parameters. Because of this, it is necessary for models to adjust processes that build the stellar populations of massive galaxies (such as mergers and star-formation efficiency), as well as processes that suppress star-formation (such as stellar and AGN feedback and ram pressure stripping at late times).

% Don't change these lines
\bsp	% typesetting comment
\label{lastpage}
\end{document}